\documentclass[12pt,a4paper,palatino]{article}
\usepackage{epsfig}
\usepackage[tmargin=1cm,bmargin=1cm,lmargin=2cm,rmargin=2cm]{geometry}

\renewcommand{\thefootnote}{\arabic{footnote}}
\begin{document}
\pagestyle{empty}
\baselineskip 22pt
\vspace*{-1in}
\renewcommand{\thefootnote}{\fnsymbol{footnote}}
\begin{flushright}
SINP/TNP/06-08\\
{\tt hep-ph/0604135}
\end{flushright}
\vskip 65pt
\begin{center}
{\Large \bf PDF and scale uncertainties of various DY distributions in ADD 
and RS models at hadron colliders}\\
\vspace{8mm}
{\bf
M. C. Kumar$^{a,b}$
\footnote{mc.kumar@saha.ac.in}, 
Prakash Mathews$^a$
\footnote{prakash.mathews@saha.ac.in}, 
V. Ravindran$^c$
\footnote{ravindra@mri.ernet.in}
}\\
\end{center}
\vspace{10pt}
\begin{flushleft}
{\it
a) 
Saha Institute of Nuclear Physics, 1/AF Bidhan Nagar,
Kolkata 700 064, India.\\

b)
School of Physics, University of Hyderabad, Hyderabad 500 046, India.\\

c) Harish-Chandra Research Institute,
 Chhatnag Road, Jhunsi, Allahabad, India.\\
}
\end{flushleft}

\vspace{10pt}
\begin{center}
{\bf ABSTRACT}
\end{center}
\vskip12pt
In the extra dimension models of ADD and RS we study the dependence of the 
various parton distribution functions on observable of Drell-Yan process 
to NLO in QCD at LHC and Tevatron energies.  Uncertainties at LHC due to
factorisation scales in going from leading to next-to-leading order in QCD 
for the various distributions get reduced by about 2.75 times for a
$\mu_F$ range $0.5 ~Q < \mu_F < 1.5 ~Q$.  
Further uncertainties arising from the error on experimental data are
estimated using the MRST parton distribution functions.  

\vfill
\clearpage

\setcounter{page}{1}
\pagestyle{plain}

\section{Introduction}

The gauge hierarchy problem has been one of the main motivations for physics 
beyond the Standard Model (SM).  One of the directions that has emerged, 
looks at why gravity appears weak as compared to the other three 
interactions of the SM.  This apparent weakness has been accounted for 
by the existence of either large extra spatial dimensions ADD model 
\cite{add} or warped extra dimension RS model \cite{rs}.  In either case the 
fundamental Planck scale could be of the order of a TeV and hence a 
possible explanation of the hierarchy.  In both these models only gravity 
is allowed to propagate the extra dimensions while the SM particles are 
constrained on a 3-brane.  This consequently leads to Kaluza-Klein (KK) modes 
in 4-dimensions, which in ADD and RS models have distinct KK spectrum and 
their effective interaction with the SM model particles.  The experimental 
signatures of these KK modes have been of intense phenomenological activity. 
With the closure of LEP and the advent of LHC the focus now shifts to
hadron colliders.

At hadron colliders, it is important to have a precise knowledge of the 
parton distributions functions (PDFs) to predict production cross sections 
of both signals and backgrounds.  These universal  PDFs are non-perturbative 
inputs that are extracted from global fits to available data on deep-inelastic 
scattering (DIS), Drell-Yan (DY) and other hadronic processes.  They describe 
the momentum distribution of the partons in a proton and various groups have 
parametrised the PDFs for a wide range of proton momentum fraction $x$ 
carried by the parton and for the center-of-mass energy $Q^2$ at which 
the process takes place.  Parametrisation of PDFs to a particular order 
in QCD would involve various theoretical and experimental uncertainties.
Recently there has been a series of papers \cite{us}-\cite{us3} which for the 
first time have calculated the next-to-leading order (NLO) QCD corrections 
to various distributions of DY process for both ADD and RS model. 
In \cite{us1}, NLO-QCD corrections to dilepton production at 
hadron colliders in the ADD model were presented for the first time, 
this was extended to the RS model in \cite{us2}.  Further in \cite{us3} 
we have considered the double differential cross section, 
$d^2\sigma/dQ^2/d\cos\theta^*$, for dilepton production in models with 
large and warped extra dimensions.  The $\cos\theta^*$ distribution is the 
one that is actually used by experiments and hence is of particular 
importance.  These NLO results would certainly reduce one aspect of the 
theoretical uncertainties as results prior to this calculation were only 
at leading order (LO) in QCD for process involving gravity.

In \cite{us1,
us2,us3} we have quantified the theoretical uncertainties coming from the
QCD corrections by computing all the processes that enter at NLO level. 
Unlike the standard model contribution to DY, the extra-dimensional models 
bring in more processes even at the LO level.  For example, in 
these models, the gluon initiated process enter at the LO level 
in addition to quark anti-quark initiated process.  At LHC, the gluon initiated 
process is more sensitive to factorisation scale compared to quark initiated 
process that necessitated the relevance of NLO computation.  It was found in 
\cite{us1,us2,us3} that the NLO corrections are considerably large and the 
factorisation scale uncertainty goes down significantly with these 
corrections as expected.  Our entire analysis is model independent
because the QCD corrections factor out from the model dependent quantities.
In \cite{us1,us2,us3}, we used the MRST parton density sets.  It is well
known that the different PDF sets themselves can affect the theoretical 
predictions and it is important to quantify these effects in the observable 
that could probe new physics.  With this in mind we have performed a model 
independent analysis on uncertainties coming from the choice of PDF sets 
in order to make our predictions more reliable.
In this paper we have looked at the dependence due to 
various PDFs for the production of dilepton at LHC and Tevatron 
including gravity effects in the ADD and RS models incorporating 
the NLO QCD corrections.  The PDF sets used in this study are Alekhin 
\cite{alekhin}, CTEQ \cite{cteq02} and MRST \cite{mrst01}.  Dependence 
on PDF sets is also compared with experimental errors
that enter the parametrisation of the PDF, which are now available to 
NLO QCD \cite{err0,err1}.  For this purpose we used the MRST distribution 
\cite{err1} as 
a typical case.  The dependence of factorisation scale $\mu_F$ and 
renormalisation scale $\mu_R$ in going from LO to NLO is also studied.

For the ADD model we also study the dependence on UV cutoff of the KK mode 
sum by keeping the UV cutoff different from the scale of the model $M_S$.
This prescription allows us to study the cutoff dependence.  The dependence
of the cross section on the number of extra dimension as a result of this 
prescription is similar to that of the real graviton production case.  The 
ADD model is a low energy effective theory valid below the scale $M_S$, it is 
conventional to equate the cutoff to the scale of the effective theory 
\cite{grw,hlz}.

The plan of the paper is as follows.  In section 2 we describe the extra
dimension models studied and describe the parameters of the models.  In 
section 3 we discuss the theoretical uncertainties {\em viz.} the PDF
uncertainties, renormalisation and factorisation scale dependences.  The
improvement of the scale dependence in going from LO to NLO is also 
discussed.  In section 4 we look at the dependence due to the error on 
the data and estimate the experimental error for a few observable using 
the MRST PDF.  Finally we summarise our results in section 5.

\section{Extra Dimension Models}

Extra dimension models that allow gravity to propagate the extra dimensions
would in 4-dimensions have KK modes which couple to SM particles through the
energy momentum tensor.  The Feynman rules of the KK mode interactions with 
the SM fields are given in \cite{grw,hlz}.  Due to the different methods of 
compactification of the extra dimensions in ADD and RS models, their KK 
spectrum are very distinct.  Experimental signature of extra dimensions would 
correspond to deviation from SM predictions due to the virtual exchange of 
KK modes or direct production of KK modes at a collider.

In the ADD case, there is a tower of KK modes which are almost degenerate in 
energy and a sum over these KK modes gives an observable effect.  In the 
case of dilepon production, in addition to the SM photon and $Z$ production 
modes, one has to take into account virtual KK modes.  Performing the sum 
over the virtual KK modes leads to an integral which has to be regulated by 
an UV cutoff.  The propagator after the KK mode summation becomes 
\begin{eqnarray}
\kappa^2 {\cal D} (Q^2) \equiv \kappa^2 \sum_n \frac{1}{Q^2-m_n^2+i 
\epsilon}= \frac{16 \pi}{M_S^4} \Big(\frac{Q}{M_S}\Big)^{d-2} 
I\Big(\frac{\Lambda_c}{Q}\Big) ~,
\label{prop}
\end{eqnarray}
where $\kappa=\sqrt{16 \pi}/M_P$ is the strength of the gravitational 
coupling to the SM particles, $m_n$ the mass of KK modes, $d$ is the 
number of extra dimensions and $M_S$ is the scale of the $4+d$ 
dimensional theory.  The summation over the non-resonant KK modes yields 
$I(\Lambda_c/Q)$ \cite{hlz}.  Conventionally the UV cutoff $\Lambda_c$ 
is identified with the scale of the extra dimension theory $M_S$, which 
simplifies the expression giving a mild dependence on the number of extra 
dimensions \cite{grw,hlz}.

In this analysis, we have kept the cutoff $\Lambda_c$ different from $M_S$
\footnote{Effects of the various UV cutoff methods on the low scale quantum 
gravity model have been discussed in \cite{gs}.}.
Note that the summation of KK modes in Eq.~(\ref{prop}), modifies the $M_P$ 
suppression to $M_S$ suppression.  The ADD model is a effective low energy 
theory valid below the scale $M_S$, for consistency it is essential to 
satisfy the condition $Q<\Lambda_c<M_S$.  The parameters of the ADD model 
are $M_S$ the scale of the $4+d$ dimensional theory and $d$ the number of 
extra spatial dimensions.  If $\Lambda_c \ne M_S$ then there is an 
additional parameter.  We have studied the dependence of the cross section 
on the cutoff $\Lambda_c=\alpha M_S$ and varied $\alpha=0.7-1$.  In 
Fig.~\ref{lambda}a we see that the cross section decreases as we lower the 
cutoff $\Lambda_c$.  The corresponding K-factor also decreases for lower 
cutoff Fig.~\ref{lambda}b.
Dependence of the cross section on the number of extra dimensions $d$ is 
shown in Fig.~\ref{lambda}c for $\Lambda_c=M_S$, the cross section decreases 
as $d$ increases.  Reducing $\Lambda_c$ decreases the cross sections and
if $d$ is increased it brings down the cross section much faster
Fig.~\ref{lambda}d.  Large extra dimension searches in the dimuon channel at
the Tevatron \cite{d01} have put bounds on $M_S$ in the range 0.8 - 1.27 
TeV.  For the analysis, they have used the double differential cross section 
with respect to invariant mass and the $\cos \theta^*$ \cite{cl}.

In the RS model, the gravity propagate one extra dimension which is warped 
by an exponential factor $\exp(-\pi k L)$, where $L$ is the compactification 
length and $k$ is the curvature of the $AdS_5$ space-time.  The parameters 
of the RS model are $m_0= k \exp(-\pi k L)$ which sets the mass scale of 
the KK modes and $c_0=k/M_P$ the effective coupling.  The higher KK modes 
have enhanced coupling to SM 
particles due to the warp factor and decouple from the zero mode, which 
is as usual $M_P$ suppressed.  RS KK spectrum is distinct from the ADD
case and hence the summation of the KK modes that contribute to the virtual
process would also be different.  The function ${\cal D}(Q^2)$ in the KK mode
propagator results from summing over the resonant KK modes and is given by
\begin{eqnarray}
{\cal D}(Q^2) &=& \sum_{n=1}^\infty \frac{1}{Q^2 - M_n^2 + i M_n \Gamma_n}
\equiv {\lambda \over m_0^2} \ ,
\end{eqnarray}
where $M_n$ are the masses of the individual resonances and $\Gamma_n$
are the corresponding decay widths.  The graviton widths are obtained by 
calculating their decays into final states involving SM particles.  $\lambda$ 
is defined as
\begin{eqnarray}
\lambda (x_s) & = & \sum_{n=1}^\infty
\frac{x_s^2 -x_n^2 -i \frac{\Gamma_n}{m_0} x_n}
       {x_s^2 -x_n^2 +  \frac{\Gamma_n}{m_0} x_n} \ ,
\end{eqnarray}
where $x_s=Q/m_0$.  We have to sum over all the resonances to get the value
of $\lambda(x_s)$, which is done numerically for a given value of $x_s$.
Searches for the RS KK modes at Tevatron in the dielectron, dimuon and digamma
channel \cite{d02} have yielded a lower limit between 250 - 785 GeV depending 
on the coupling to the SM particles.

\section{Theoretical uncertainties}

In the QCD improved parton model the hadronic cross section can be expressed
in terms of pertubatively calculable partonic cross sections denoted by
$\hat \sigma^{ab} (\tau, Q^2, \mu_F)$
convoluted with appropriate non perturbative partonic flux  $\Phi_{ab}(\tau,\mu_F)$
at a factorisation scale $\mu_F$. 
The subprocess cross section is a
perturbative expansion in the strong coupling constant $\alpha_s (\mu_R)$
and is calculated order by order in $\alpha_s$.
Here $\mu_R$ is the renormalisation scale and $\tau=Q^2/S$ is the DY scaling 
variable.  In perturbative QCD, the unknown higher order corrections
and the scale uncertainties are strongly correlated.
The factorisation of mass singularities from the perturbatively
calculable partonic cross sections leads to the introduction
of factorisation scale $\mu_F$ in both non-perturbative
partonic flux  $\Phi_{ab}(\mu_F)$ as well as the finite
partonic cross sections $d\hat \sigma_{ab}(x,\mu_F)$. 
Even though the choice of the scale is guided
by the hard scale of the problem, the exact value does not
come from the theory. 
The PDFs and partonic cross sections satisfy renormalisation group
equations such that the hadronic cross section is independent of
the factorisation scale $\mu_F$. 
In addition to the factorisation scale, the partonic cross sections
are dependent on the renormalisation scale $\mu_R$. 
The choice of the scale is again arbitrary.
Even though this is an advantage to choose appropriately to do
perturbative calculations, it also introduces theoretical
uncertainties through the size of unknown higher order corrections.
Usually, one chooses this scale such that the perturbative methods can be
applied and then computes higher order corrections sufficiently such that
the exact choice of this scale becomes almost immaterial. 
Gravity couples to the SM fields via its energy momentum tensor, and the 
calculations are done in the high energy limits where masses of the SM 
particles are ignored.  Only parameter that requires UV renormalisation 
is the strong coupling constant, because of this we have the following 
expansion for the mass factorised partonic cross section:
\begin{eqnarray}
d\hat \sigma_{ab}(z,\mu_F^2)=\sum_{i=0}^\infty a_s^i(\mu_R^2) d\hat 
\sigma_{ab}^{(i)} (z,\mu_F^2,\mu_R^2) \ ,
\end{eqnarray}
where the coupling constant satisfies standard renormalisation group equation.
Since we are only interested in the NLO order corrections,
the Altarelli-Parisi kernels $P^{(0)}(z)$, $P^{(1)}(z)$ and the coefficients 
$\beta_0,\beta_1$ are sufficient for our analysis.
The scale uncertainties come about from the truncation of
the perturbative series. 
Unlike the perturbatively calculable partonic cross sections,
the PDFs being non-perturbative in nature are extracted from various
experiments.  These are fitted at a scale of the experiments and then
evolved according to the AP evolution equations to any other relevant
scale.  They are not only sensitive to experimental errors but also
to theoretical uncertainties that enter through the partonic
cross section calculations and the splitting functions that are known
only to certain orders in strong coupling constant in perturbative QCD.
Here, we mainly concentrate
on the uncertainties coming from PDFs in detail and quantify their impact
on the new physics searches in extra dimensional models.

\subsection{PDF uncertainty}
We first focus on the uncertainties coming from different PDF sets.
The parton flux factor for both LHC 
and Tevatron would give an idea as to which component would be dominant
in the kinematical region of interest.  This flux factor enters the 
cross section.  The gluon flux is clearly
much larger in the kinematical region of interest at LHC and for
Tevatron the $q \bar q$ flux is the dominant contribution.

In the context of extra dimension theories we consider the dilepton 
production at LHC and Tevatron for both large extra dimension and warped
extra dimension models.  The process of interest is $P_1(p_1) + P_2(p_2) 
\to \mu^+(l_1) + \mu^- (l_2) + X(P_X)$, where $P_1$ and $P_2$ are the incoming 
hadrons, $\mu^\pm$ are the final lepton pair and $X$ the final inclusive 
hadronic state.  The dilepton in these models could also be produced 
from the exchange of a KK mode in addition to the usual SM gauge boson 
exchange.  Hence at LO itself the $gg$ subprocess
could contribute to the dilepton production via a KK mode exchange
in addition to the $q \bar q$ subprocess.

For both new physics searches and precision SM physics at hadron colliders 
it is essential to understand the uncertainties associated with PDFs.  We 
essentially study to what extent the cross sections depend on the various 
PDFs {\em viz.} Alekhin \cite{alekhin}, CTEQ \cite{cteq02}
and MRST \cite{mrst01}.  In the Table~1, we have tabulated the 
particular PDF that is chosen for the study and also the corresponding 
$\Lambda_{QCD}$ parameter that is used to determine the strong coupling 
$\alpha_s$. 
\footnote{In the case of Alekhin the PDF itself generates the value 
of $\alpha_s$ and is hence not tabulated.}
\vspace{.3cm}
\begin{table}[h!b!p!]
\begin{center}
\begin{tabular}{|l|l||l|l|}
\hline
\multicolumn{2}{|c||}{LO}&\multicolumn{2}{c|}{NLO}\\
\hline
PDF & $\Lambda_{QCD}$ {(GeV)} &PDF& $\Lambda_{QCD}$
{(GeV)}\\
\hline\hline
MRST2001 LO& 0.220&MRST2001 NLO&0.323\\
CTEQ6L&0.326&CTEQ6M&0.326\\
\hline
\end{tabular}
\end{center}
\label{tab1}
\caption{The PDF set used in the analysis along with the respective 
$\Lambda_{QCD}$.}
\end{table}

These groups perform a global analysis of a wide range of DIS and other
scattering data to get best fits to a particular order in QCD.  Though
all these parametrisation satisfies the general constraints, they could
differ from each other.  This is expected as PDFs are not by itself 
physical quantities and are extracted subject to experimental and theoretical
uncertainties and various assumptions and initial conditions used by the 
different groups. 
Differences among the various PDFs would translate as uncertainties on 
the physical observable.

To NLO in QCD for various PDFs, we now present 
the comparison plots for the following differential distributions
\begin{eqnarray}
\frac{d \sigma}{dQ}~, \qquad \qquad  \frac{d^2 \sigma}{dQ ~d Y}~, 
\qquad \qquad \frac{d^2 \sigma}{dQ ~d \cos \theta^*}~.
\label{dist}
\end{eqnarray}
We would look at the invariant mass distribution $Q$, the double differential
cross section with respect to $Q$ and rapidity $Y$ and the double differential
cross section with respect to $Q$ and $\cos \theta^*$.  The angle $\theta^*$
is the angle between the final state lepton momenta and the initial state
hadron momenta in the {\em c.o.m} frame of the lepton pair.  The corresponding 
K-factor which is the ratio of NLO to LO of the above distributions are also 
plotted for the various PDFs.  For the double differential cross section we 
fix the invariant mass $Q$ in the region of interest of extra dimensions 
and plot the cross section with respect to rapidity $Y$ and $\cos \theta^*$.
The first two distributions in Eq.~({\ref{dist}) are $\cos \theta^*$ integrated
distributions and hence are independent of the interference between the SM
background and the low scale gravity effects \cite{us1}.  The double 
differential with respect to $Q$ and $\cos \theta^*$ would contain the 
interference terms, but numerically it is not very significant \cite{us3}. 
Consequently even for the $\cos \theta^*$ distributions we can express
the $K$ factor of the model involving both SM and gravity as
\begin{eqnarray}
K^{(SM+GR)}(Q) = \frac{K^{SM} + K^{GR} K^{(0)}}{1+K^{(0)}} \ ,
\label{eq18}
\end{eqnarray}
where $K^{GR}$ is the $K$ factor of the pure gravity part.  We have 
introduced a quantity $K^{(0)}$, defined as the ratio of the LO 
distribution of gravity to SM, given by
\begin{eqnarray}
K^{(0)}(Q)=\Bigg[ {d \sigma_{LO}^{SM}(Q) \over dQ} \Bigg]^{-1}
        \Bigg[ {d \sigma_{LO}^{GR}(Q) \over dQ} \Bigg] \ .
\label{eq19}
\end{eqnarray}
The behaviour of $K^{(0)}(Q)$ is governed by the competing coupling constants
of SM and gravity and the parton fluxes involved.  Basically the factor 
$K^{(0)}$ is an indicator as to the source of the total $K^{(SM+GR)}$-factor.
$K^{(0)}(Q)$ as a function of $Q$ rises much faster for LHC than Tevatron and 
reaches 1 much earlier.  Since the $gg$ subprocess contributes at LO
itself for the gravity mediated process, the gravity effects are much
larger at the LHC where the gluon flux is much larger.  This would also 
result in larger K-factor for the process at LHC at large Q where the
gravity contribution dominates.  At Tevatron since the gluon flux is 
smaller the K-factor is similar to the SM K-factor.
For both ADD and RS models the signal for new physics is the excess of
events in the total cross section or the various distribution over the
SM background.  If we restrict ourself to these extra dimensional
models, the signal is due to the effect of the KK modes and can not be
mimicked by the SM.  We would like to emphasise that we are not analysing 
the existing Tevatron data to extract bounds on the ADD and RS parameters, 
which would need a full hadron-level simulation, but estimate the various 
uncertainties to NLO in QCD by choosing typical representative values 
for the ADD and RS parameters.

We begin with the ADD model wherein we have chosen $d=3$ and $M_S=2$ TeV.  In 
Fig.~\ref{invQ}a the cross section is plotted as a function of the invariant 
mass $Q$ of the dilepton at LHC for the various PDFs.   There is only a mild
dependence on the difference in the PDFs, but when plotted for the corresponding
K-factor then the PDF dependence is larger for both low and high values of $Q$,
Fig.~\ref{invQ}b.  At low $Q$ it is the SM part which contributes to the 
K-factor while at high $Q$ it is the beyond SM effects that contribute.
At low $Q$ where the K-factor is due to SM part, MRST and CTEQ 
are similar, while Alekhin is smaller.  At large $Q$ the 
K-factor is due to the gravity part and here CTEQ is larger.

For the double differential cross section with respect to invariant mass 
distribution and rapidity $Y$ Fig.~\ref{rapY}a, we have plotted as a function
of rapidity $Y$ for a fixed $Q=0.7$ TeV.  Only in the central rapidity
region do the PDFs differ, with MRST being the dominant while CTEQ  is the 
smallest.  The K-factor is quite large at the central rapidity region and 
would range from 1.5 - 1.6 depending on the PDF used.  The general
behaviour of the K-factor is similar for MRST and Alekhin.  At large 
rapidities $y=\pm 2$ the K-factors are quite different with Alekhin being 
1.25 while CTEQ the largest is 1.45.  For $Q=0.7$ TeV the K factor is large
which we can see from Fig.~\ref{invQ}b, wherein the dominant contribution is 
from the gravity mediated $gg$ initiated subprocess.  

In Fig.~\ref{rapY}c we have plotted the double differential cross section with
respect to $Q$ and $\cos \theta^*$ as a function of $\cos \theta^*$ for a fixed
$Q=0.7$ TeV.  MRST gives the largest and CTEQ the least with Alekhin being
a central value in the spread.  The difference exists for the full range of 
$\cos \theta^*$.  
The SM background has a different $\cos \theta^*$ dependence. The 
interference of the SM and the gravity effect is not zero for 
the $\cos \theta^*$ distribution but does not contribute significantly.
The K-factor for central $\cos \theta^*=0$ region is about 1.52 but differ 
with PDFs as $\cos \theta^* \to \pm 1$, Fig.~\ref{rapY}d.  
Since there is no $gg$ initiated process in the SM background to NLO the 
K-factor is much smaller.

In the RS model we have chosen the mass of the first KK mode $M_1=1.5$ TeV 
and the coupling $c_0 =0.01$.  In Fig.~\ref{rsQ}a we have plotted invariant 
mass distribution of the dilepton in the RS model.  At the KK mode resonances
the cross section differs from the SM cross section, but the dependence on
the PDFs are very mild.  In Fig.~\ref{rsQ}b the corresponding K-factors are
plotted for various PDFs. 
The behaviour of the K-factor of the RS model can be understood with the help 
of Eq.~(\ref{eq18},\ref{eq19}).  It is only in the RS graviton resonances 
region that $K^{(0)}$ is large and hence the $K$-factor is dominated by the 
$K^{GR}$ factor.   In the off resonance regions it is the $K^{SM}$ which 
contributes.
There is a wide difference in the K-factor more in the second peak and even 
off peak where the effect is mainly SM.  This is due to the high $Q$ value 
that is chosen in the RS case.

For the double differential with respect to rapidity and invariant mass,
in Fig.~\ref{rsY}a we have plotted it for the rapidity range of LHC for 
$Q=1.5$ TeV, which is the region of the first RS KK mode.  It is only in
the resonance region that the effects of RS are visible.  Here 
there seems to be a clustering of PDFs but for CETQ in the central 
rapidity region.  In the central rapidity region the K-factor varies from
1.6 - 1.75 Fig.~\ref{rsY}b.  
In the first RS KK resonance region at $Q=1.5$ TeV the gravity
dominates and hence the K-factor is large (Eq.~(\ref{eq18})).
Beyond the central rapidity region $Y=0$ the 
K-factor dependence on PDFs is substantial.
In Fig.~\ref{rsY}c we have plotted the double differential with respect to 
$\cos \theta^*$ for $Q$ fixed at the first resonance.  The cross section is
largest for $\cos \theta^*=0$ and MRST is the largest among the PDFs. 
The K-factor in Fig.~\ref{rsY}d is about 1.65 for wide range of $\cos 
\theta^*$ for Alekhin and MRST but for CTEQ it varies between 1.7 - 1.8.

In the above we discussed the extra dimension effects at LHC, now we look
at the Tevatron.  For the ADD case, in Fig.~\ref{tev1}a we have plotted the 
invariant mass distribution for the various PDFs.  The spread due 
to the various PDFs over the $Q^2$ range is not too large.  Only at large 
$Q$ there is some deviation from the SM result which is plotted in 
Fig.~\ref{tev1}a.  The K-factor for the $Q$ distribution for the various 
PDFs are plotted in Fig.~\ref{tev1}b, which are in tune with the SM K-factor 
at the Tevatron.  In Fig.~\ref{tev1}c we have plotted the PDF 
comparison plot for the rapidity distribution at $Q=0.7$ TeV.  CTEQ
and MRST plots are very similar while Alekhin is larger 
in the central rapidity region.  In the $Y=0$ region, the 
K-factor for CTEQ is about 1.1 while for MRST and Alekhin it is about 1.2,
which is in the range of the SM K-factor, Fig.~\ref{tev1}d.

For the RS model the PDF comparison plots are given in Fig.~\ref{tev2}.
In Fig.~\ref{tev2}a we have the invariant mass distribution and the 
deviation from the SM is only in the resonance region.  The PDF
dependence is very mild.  
In the first resonance region the K-factor (Fig.~\ref{tev2}b) is dominated 
by $K^{GR}$ at $Q=0.7$ TeV but at Tevatron this value is not too different 
from the SM K-factor.
In Fig.~\ref{tev2}c the $\cos \theta^*$ distribution at the first 
resonance region is plotted,
CTEQ and MRST overlap while Alekhin is larger over a wide range of 
$\cos \theta^*$.  The K-factor Fig.~\ref{tev2}d is in the range of the 
SM K-factor.

\subsection{Renormalisation/Factorisation scale uncertainties}

In Fig.~\ref{reno}a we have plotted the double differential $d^2 \sigma/dQ 
dY$ in the $Y$ range for LHC energies for a fixed $Q=0.7$ TeV. 
The dependence of cross section on $\mu_R$ comes from the strong coupling 
constant at NLO and so at LO there is no $\mu_R$ dependence.  At NLO $\mu_R$ 
dependence for the $Y$ distribution is plotted for the $\mu_R$ range $0.5 ~Q 
\le \mu_R \le 1.5 ~Q$.  The $\mu_R$ spread is largest in the central rapidity 
region and would only reduce at the NNLO order level when the $\mu_R$ 
dependences would be compensated for by the dependence coming from the 
coefficient functions. 
In Fig.~\ref{reno}b we have plotted the K-factor for SM and SM+GR and see 
how it dependence on $\mu_R$.  The uncertainties due to $\mu_R$ is much 
larger when the gravity is included.  The percentage spread is of the order 
of 3.5 \% which is comparable to the $\mu_F$ spread at NLO.
\begin{table}[h!]
\begin{center}
\begin{tabular}{|c|c|c|c||c|c|}
\hline
\multicolumn{1}{|c|}{}& \multicolumn{1}{|c|}{Distributions}&
\multicolumn{2}{|c||}{Tevatron} &\multicolumn{2}{|c|}{  LHC  }\\
\cline{3-6}

{}&{} &LO& NLO& LO & NLO\\
\hline\hline
{}&{}&{}&{}&{}&{}\\
{} & $ d^{2} \sigma /dQ dY$ & 22.8 & 7.4 & 9.5&3.5\\
ADD&{}&{}&{}&{}&{}\\
{}& $ d^{2} \sigma /dQdcos \theta $ & 24.2&8.2 &10.9& 3.8\\
{}&{}&{}&{}&{}&{}\\
\hline
{}&{}&{}&{}&{}&{}\\
{} & $ d^{2} \sigma /dQ dY $ & 23.2& 7.7&18.7&6.9\\
RS&{}&{}&{}&{}&{}\\
{}& $ d^{2} \sigma / dQ d cos \theta $ & 24.2& 8.0&18.4&6.8\\
{}&{}&{}&{}&{}&{}\\
\hline
\end{tabular}
\end{center}
\caption{Percentage spread as a result of factorisation scale variation in 
the range $0.5 Q \le \mu_F \le 1.5 Q$. For the ADD case $Q=0.7$ TeV. 
For the RS first resonance region $Q=1.5$ TeV for LHC and $Q=0.7$ TeV for 
Tevatron.}
\label{table2}
\end{table}

In Fig.~\ref{fact_Y} we have plotted $Y$ distribution and its K-factor for 
ADD and RS model at a fixed $Q=\mu_R$.  The $\mu_F$ 
variation is studied by varying $\mu_F$ in the range 
$0.5 ~Q \le \mu_F \le 1.5 ~Q$.  We see that for both the ADD and RS model
in going from LO to NLO in QCD, the uncertainties due to $\mu_F$ variation 
considerably get reduced.  The spread of K-factor with $\mu_F$ is much 
smaller for the 
SM as compared to SM+GR.  This certainly indicates need to go beyond NLO. 
Similar trends are observed for the $\cos \theta^*$ distribution plotted 
in Fig.~\ref{fact_cos}.

In Table~\ref{table2} we tabulate the percentage spread of the factorisation
scale $\mu_F$ dependence in the range $0.5 Q \le \mu_F \le 1.5 Q$ for the LHC
and Tevatron.  On the average at LHC and Tevatron, the percentage spread of 
the scale variation get reduced by about 2.75 times in going from LO to NLO.

\section{Experimental Uncertainties}

In addition to the theoretical uncertainties that we have described in the 
previous section, there are uncertainties due to errors on the data. 
Various groups have studied the experimental errors and have estimates of 
the uncertainties on the PDFs within NLO QCD framework \cite{err0,err1}.
Now that NLO QCD results are also available for extra dimension searches 
\cite{us1} for the dilepton production, we consider some of the 
distributions and estimate the uncertainties due to the experimental error.
In Fig.~{\ref{err}}a we have plotted the error band for the MRST 2001 PDF
\cite{err1} in the ADD model for the dilepton invariant mass distribution 
at LHC.  This error band is comparable to the spread associated with the 
different set of PDFs as given in Fig.~\ref{invQ}a.  At $Q=1$ TeV the 
percentage of experimental error is 7.5 \% for $SM+GR$ while the pure SM 
error is about 3.3 \%.  For the RS case at LHC in the first resonance 
region at $Q=1.5$ TeV the experimental error is about 12.8 \%.
At Tevatron the ADD model experimental error
is 7.4 \% at $Q=1$ TeV.  
The experimental error 
for this distribution for the central rapidity region is about 3.5 \%
and is indicated in the Fig.~{\ref{err}}b.
In general the experimental error increases with the increase in $Q$.

\section{Conclusions}

We have studied the impact of various parton density sets at next to leading 
order in strong coupling constant $\alpha_s$ in QCD to one of the most 
important processes, namely Drell-Yan production of dileptons in hadron 
colliders such as LHC and Tevatron.  This process can probe the physics 
beyond SM through exchange of new particles that these theories predict. 
At hadron colliders, the precise measurement of DY production cross 
sections is possible.  In this context, we have studied the theories of 
extra dimensions such as ADD and RS which attempt to explain gauge 
hierarchy problem in SM.  We have discussed various theoretical uncertainties 
that enter through renormalisation, factorisation scales and the parton 
density sets.  We have quantified the uncertainties coming from various 
parton density sets using the recent results on NLO QCD 
corrections to parton level cross sections and recent PDF sets that take 
into account various theoretical and experimental errors.  Our entire 
analysis is model independent thanks to the factorisation of QCD radiative 
corrections from the model dependent contributions.  More precisely, our 
findings are independent of the finer details of the model as they factor 
out from the rest.  We find that the K-factor for various observable
depends on the choice of PDFs. 

\vspace{.3cm}
\noindent
{\bf Acknowledgment:}\\
M C Kumar, CSIR JRF would like to thank CSIR, New Delhi for financial 
support.




\begin{figure}[htb]
\centerline{
\epsfig{file=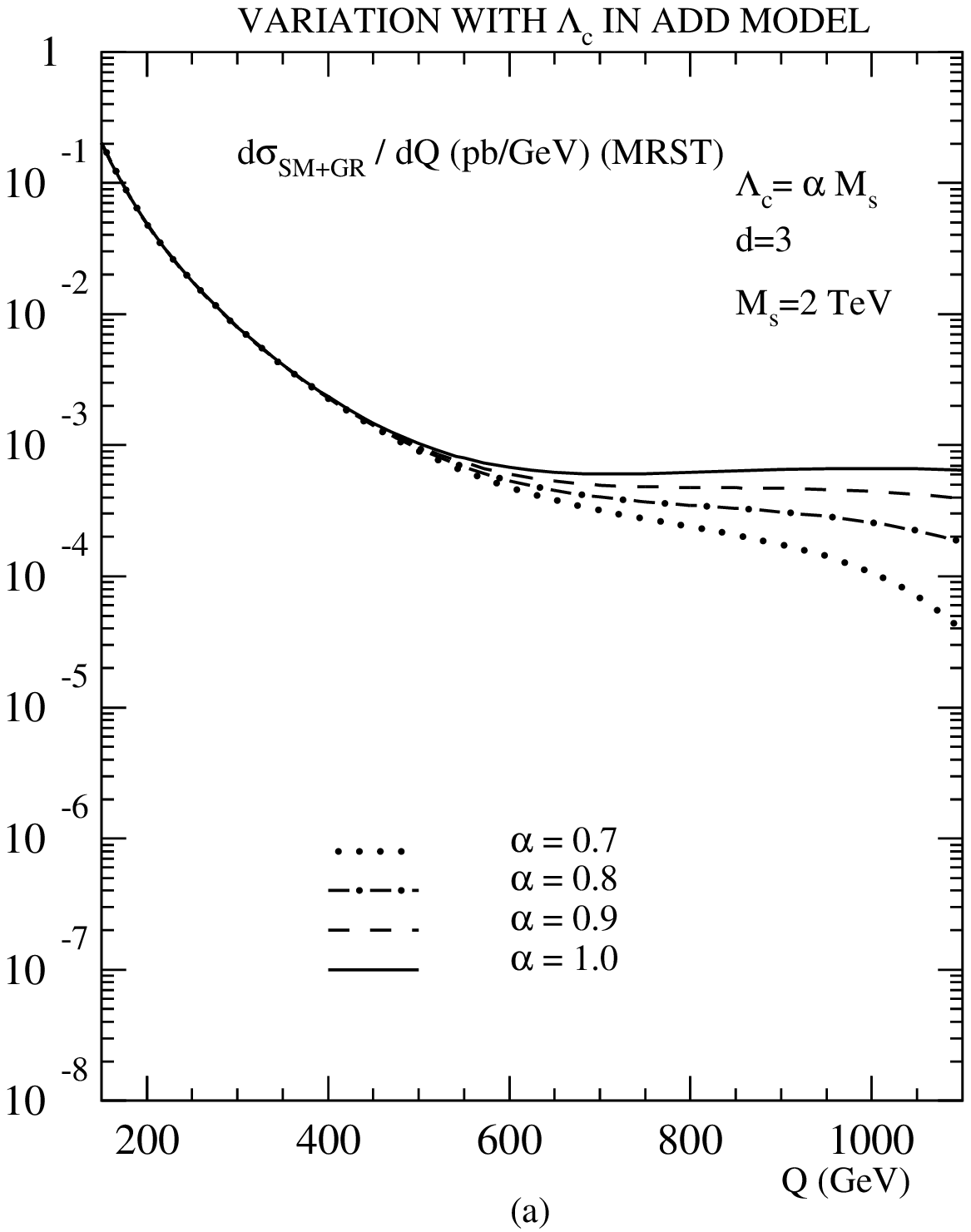,width=8cm,height=9cm,angle=0}
\epsfig{file=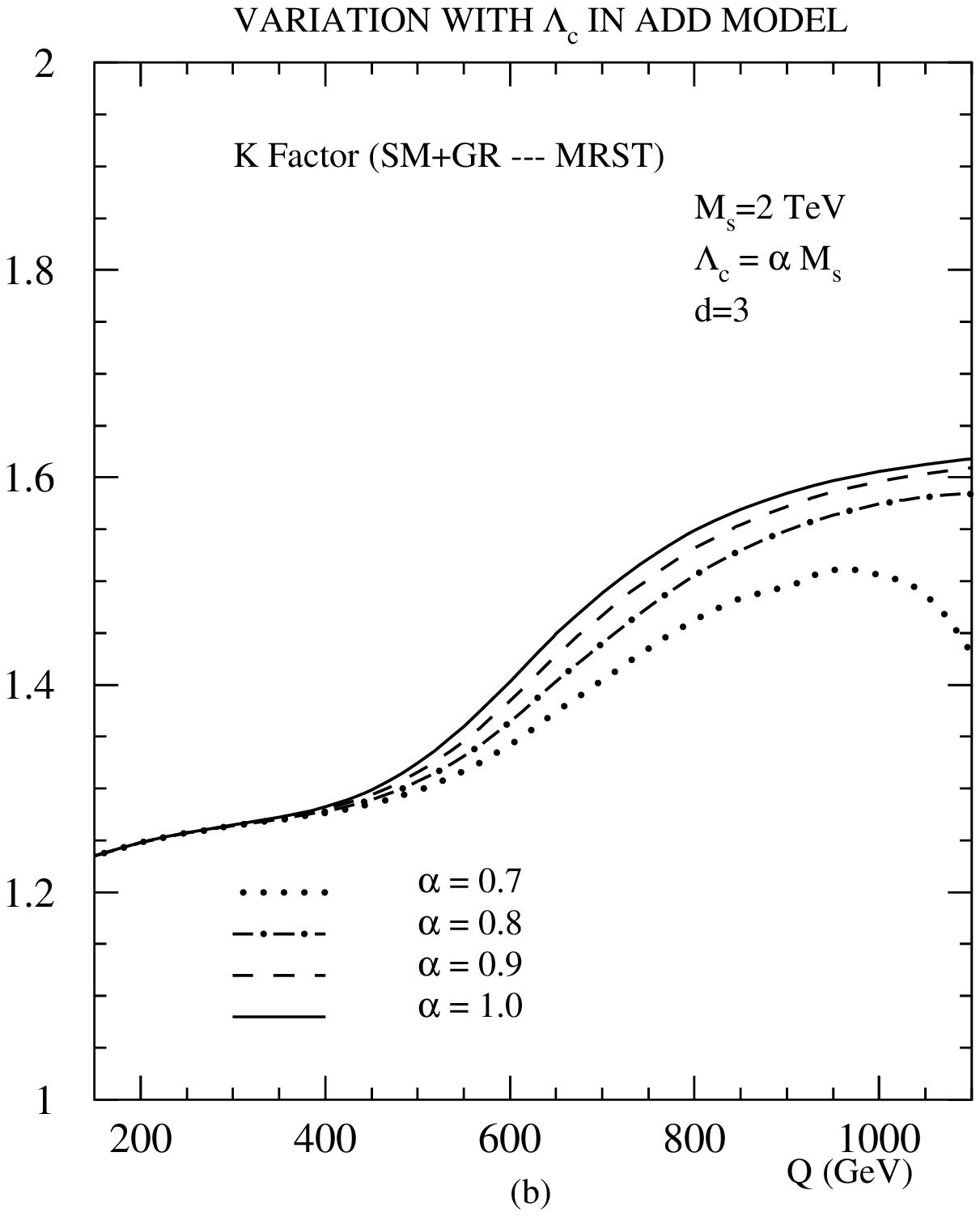,width=8cm,height=9cm,angle=0}
}
\centerline{
\epsfig{file=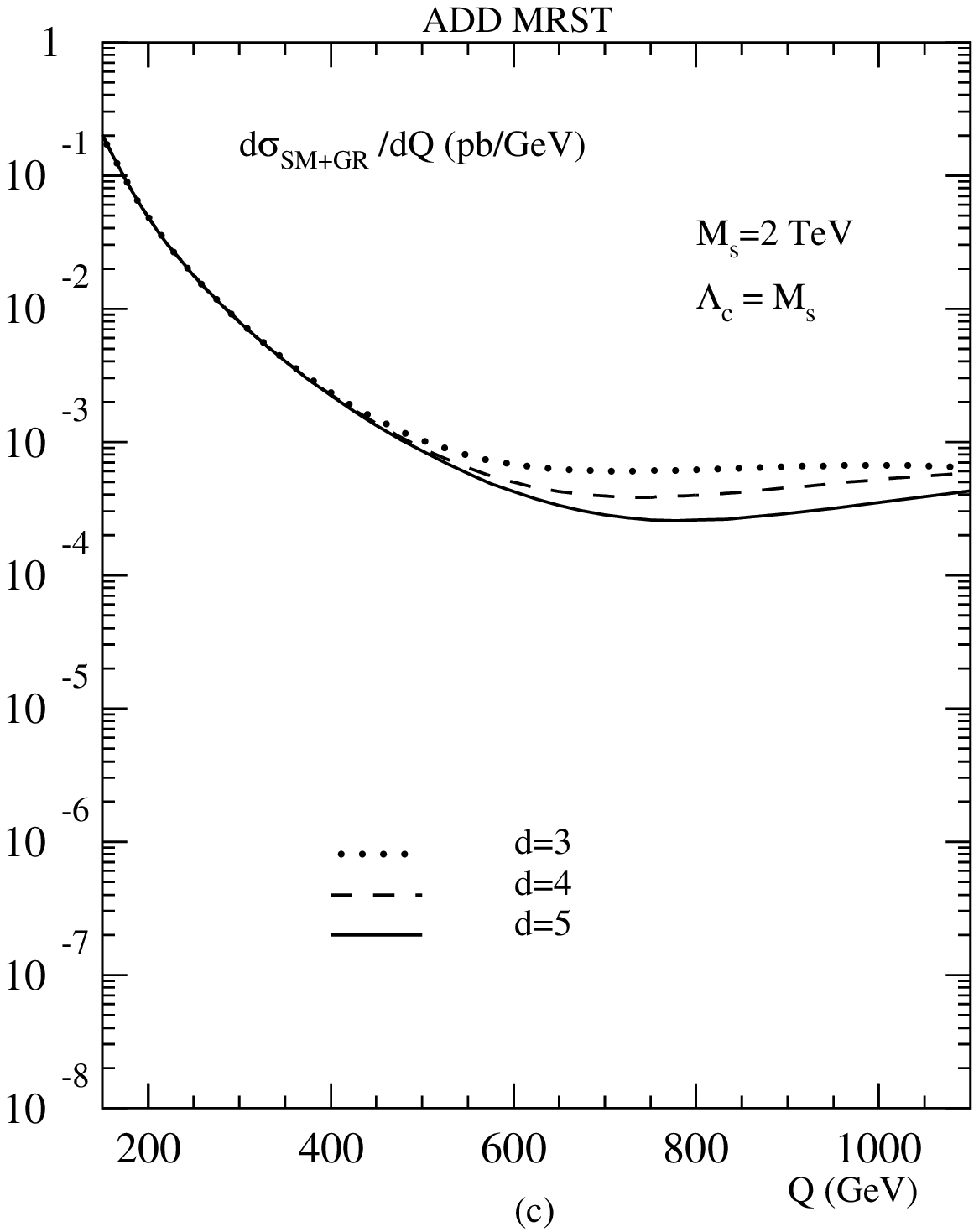,width=8cm,height=9cm,
angle=0}
\epsfig{file=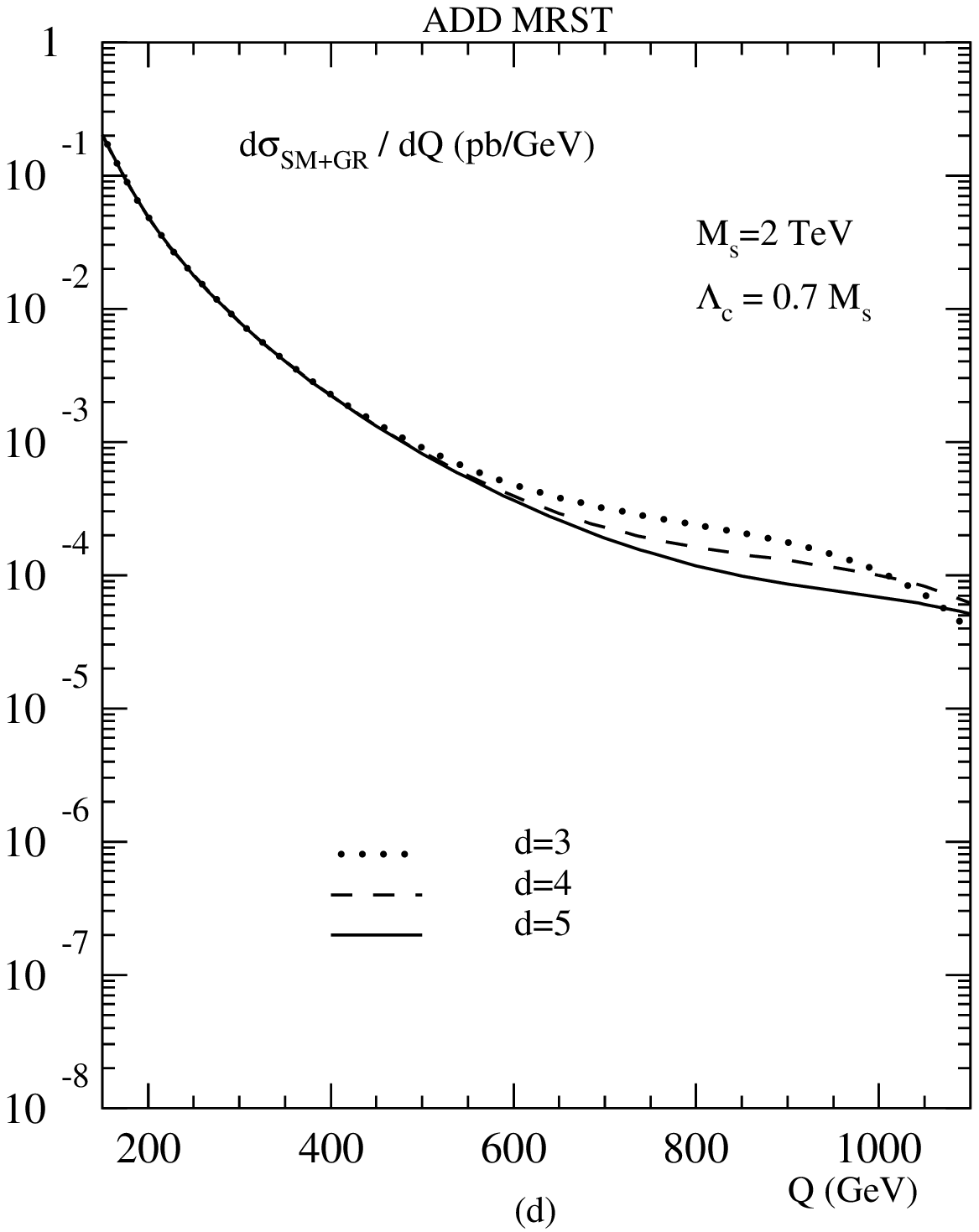,width=8cm,height=9cm,
angle=0}
}
\caption{(a) Invariant mass distribution is plotted for various values of the
cutoff $\Lambda_c=\alpha M_s$ in the ADD model. (b) The corresponding K-factor.
(c) Invariant mass distribution as a function of the number of extra spacial 
dimension $d$ for $\Lambda_c=M_S$ TeV at LHC.  (d) The same plot as (c) for 
$\Lambda_c=0.7 M_S$.}
\label{lambda}
\end{figure}

\begin{figure}[htb]
\centerline{
\epsfig{file=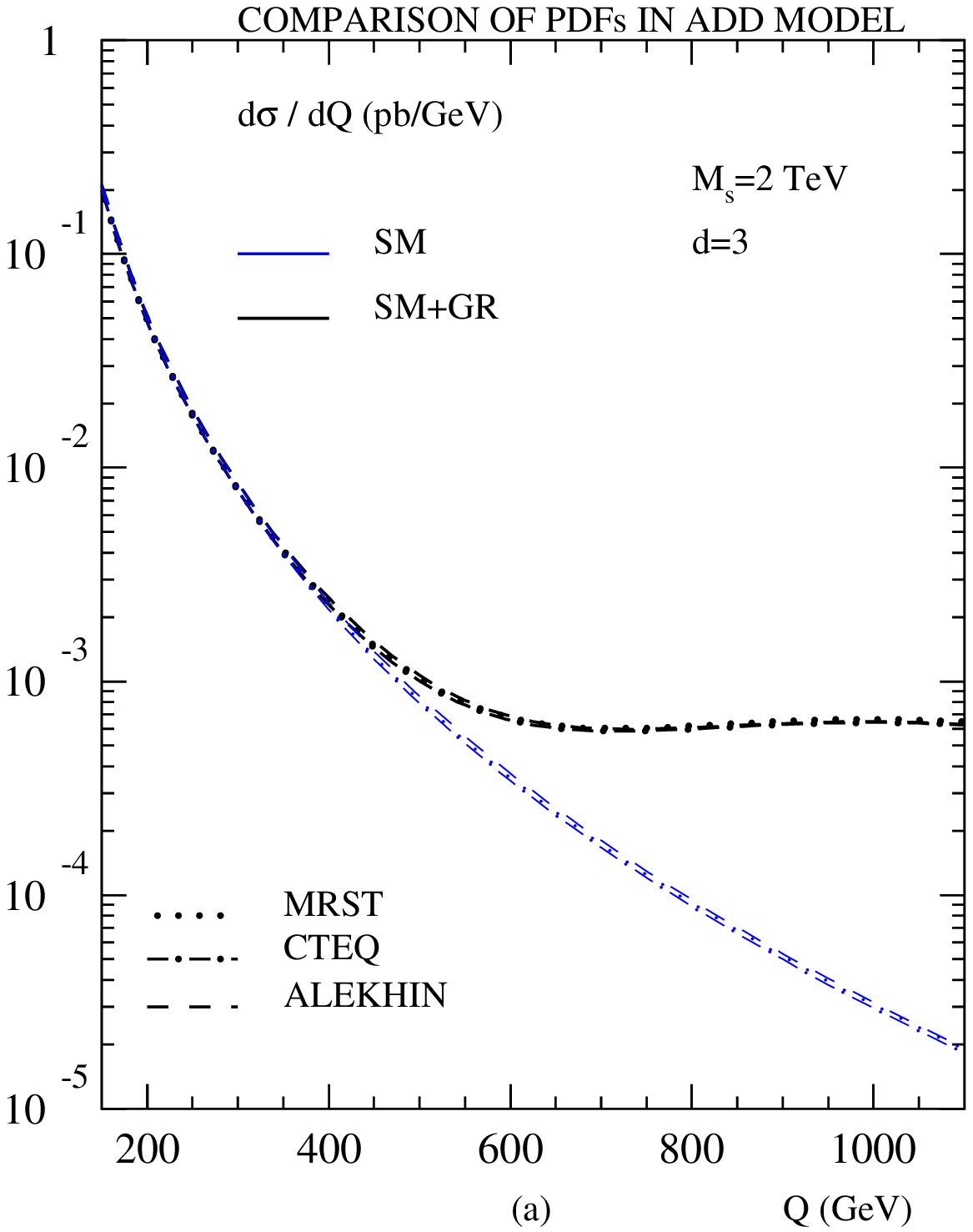,width=8cm,height=10cm,angle=0}
\epsfig{file=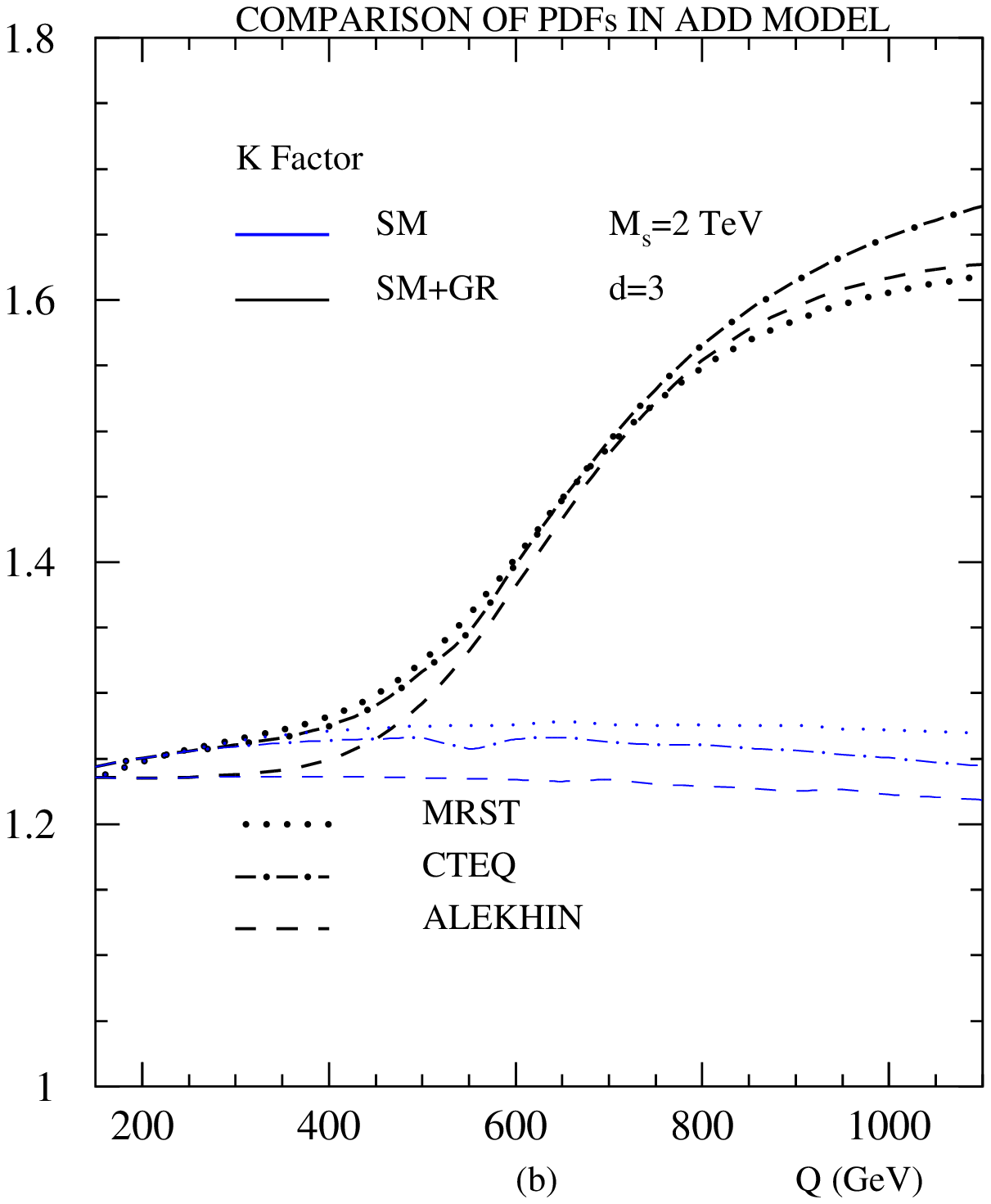,width=8cm,height=10cm,angle=0}
}
\caption{(a) Invariant mass distribution of the dilepton pair for ADD model 
with different PDFs to NLO in QCD. (b) The corresponding K-factor for the 
various PDFs.  For both (a) and (b) we have plotted the SM background to NLO
using the same line type in colour for the different PDFs.}
\label{invQ}
\end{figure}
\begin{figure}[htb]
\centerline{
\epsfig{file=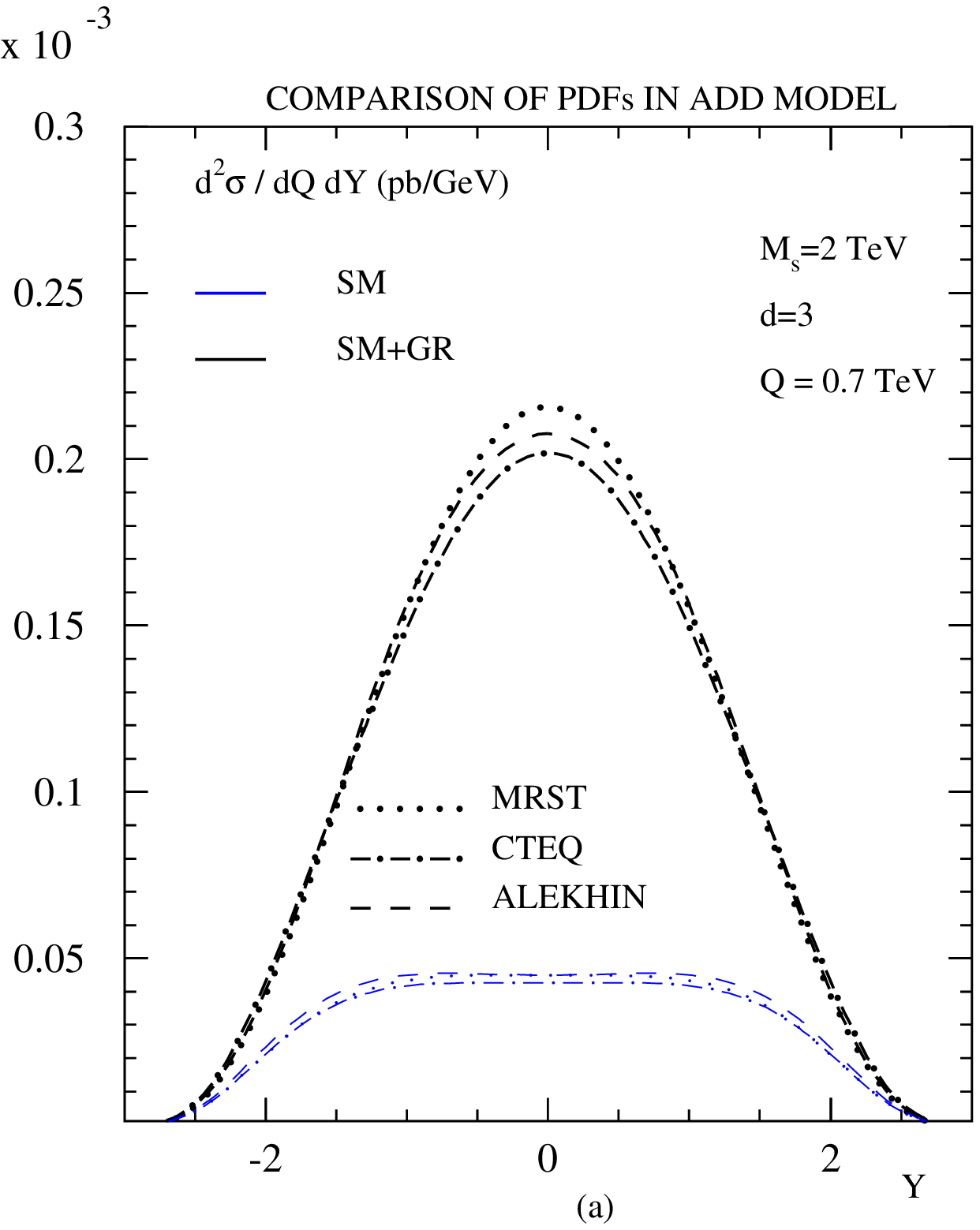,width=8cm,height=9cm,
angle=0}
\epsfig{file=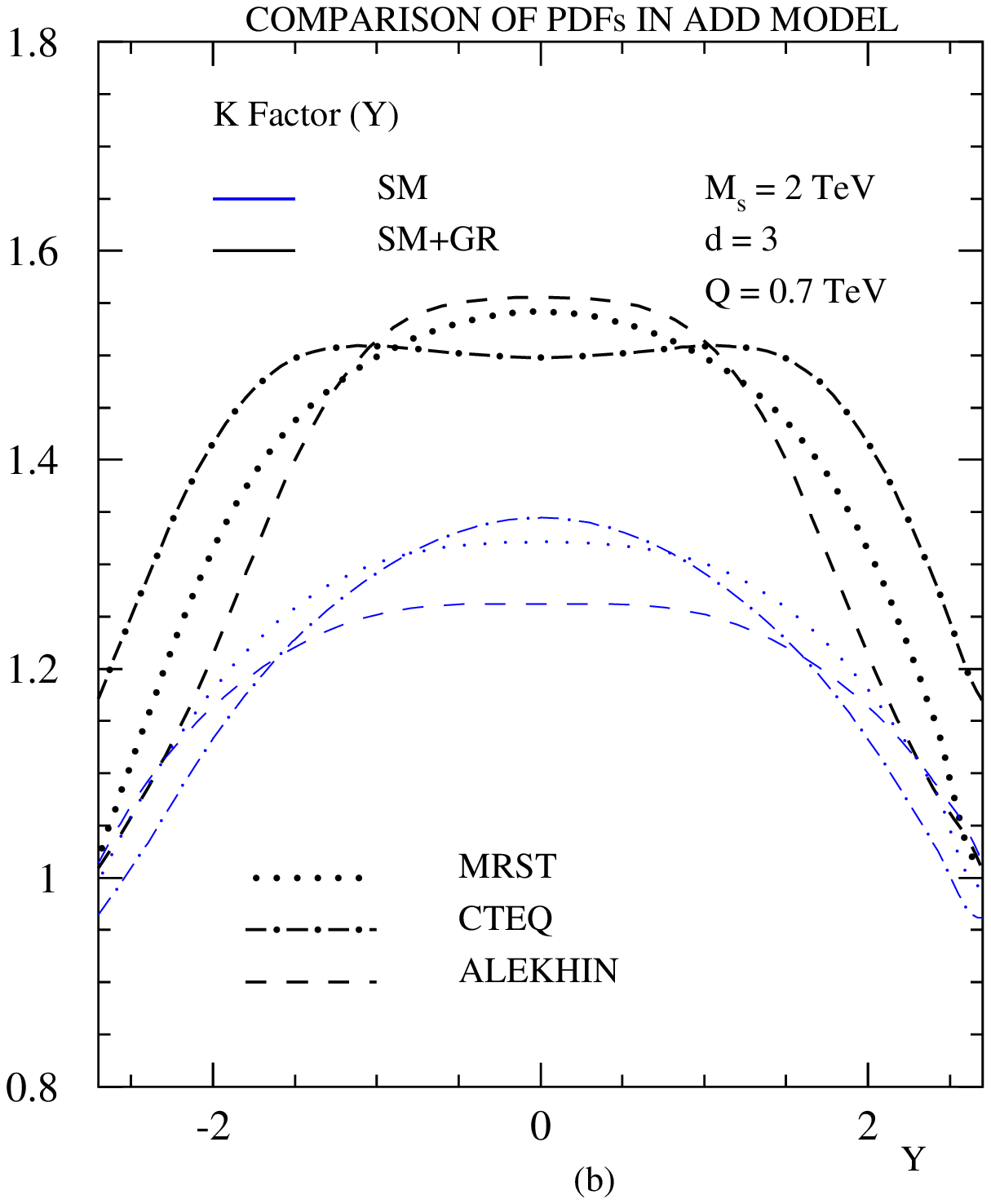,width=8cm,height=9cm,
angle=0}}
\centerline{
\epsfig{file=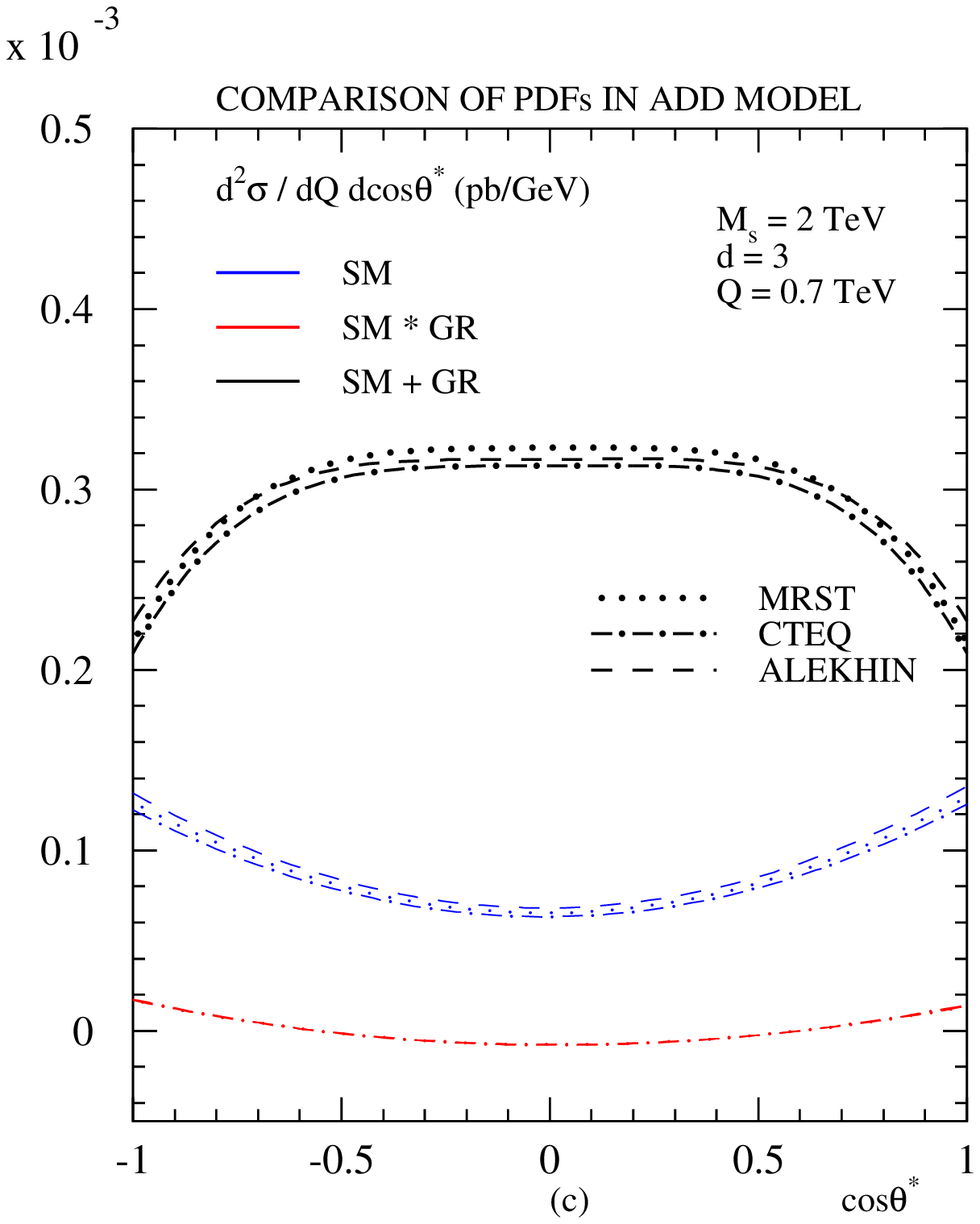,width=8cm,height=9cm,
angle=0}
\epsfig{file=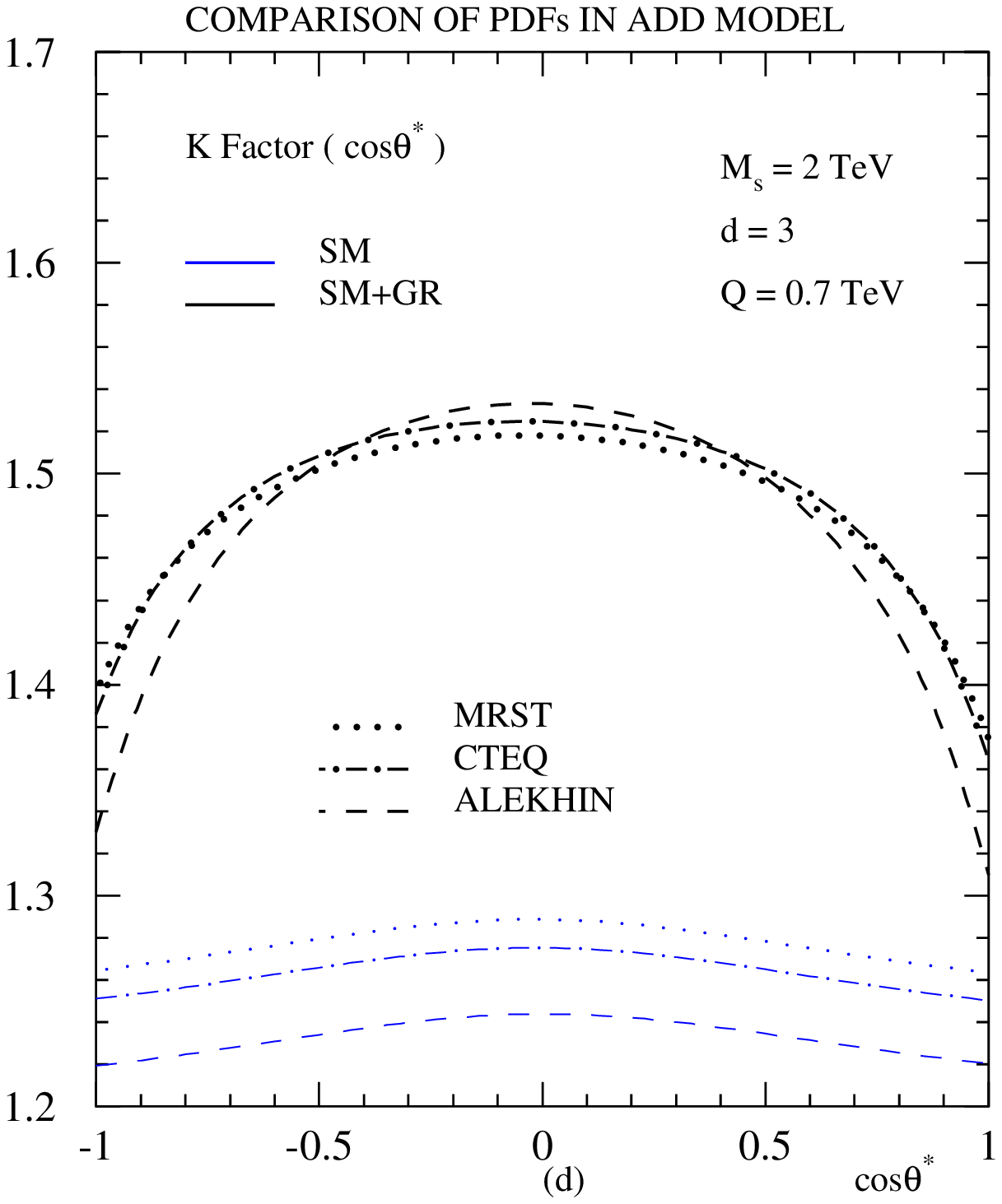,width=8cm,height=9cm,
angle=0}}
\caption{The comparison plots for the various PDFs sets for $Q=0.7$ TeV at 
LHC. (a) The double differential cross section with respect to invariant 
mass and rapidity as a function of rapidity.  
(b) The corresponding K-factor as a function of rapidity.
(c) The angular distribution of the double differential cross section
with respect to invariant mass and $\cos \theta^*$.  The interference 
of the SM background and gravity effects is also plotted.
(d) The corresponding K-factor.
}
\label{rapY}
\end{figure}

\begin{figure}[htb]
\centerline{
\epsfig{file=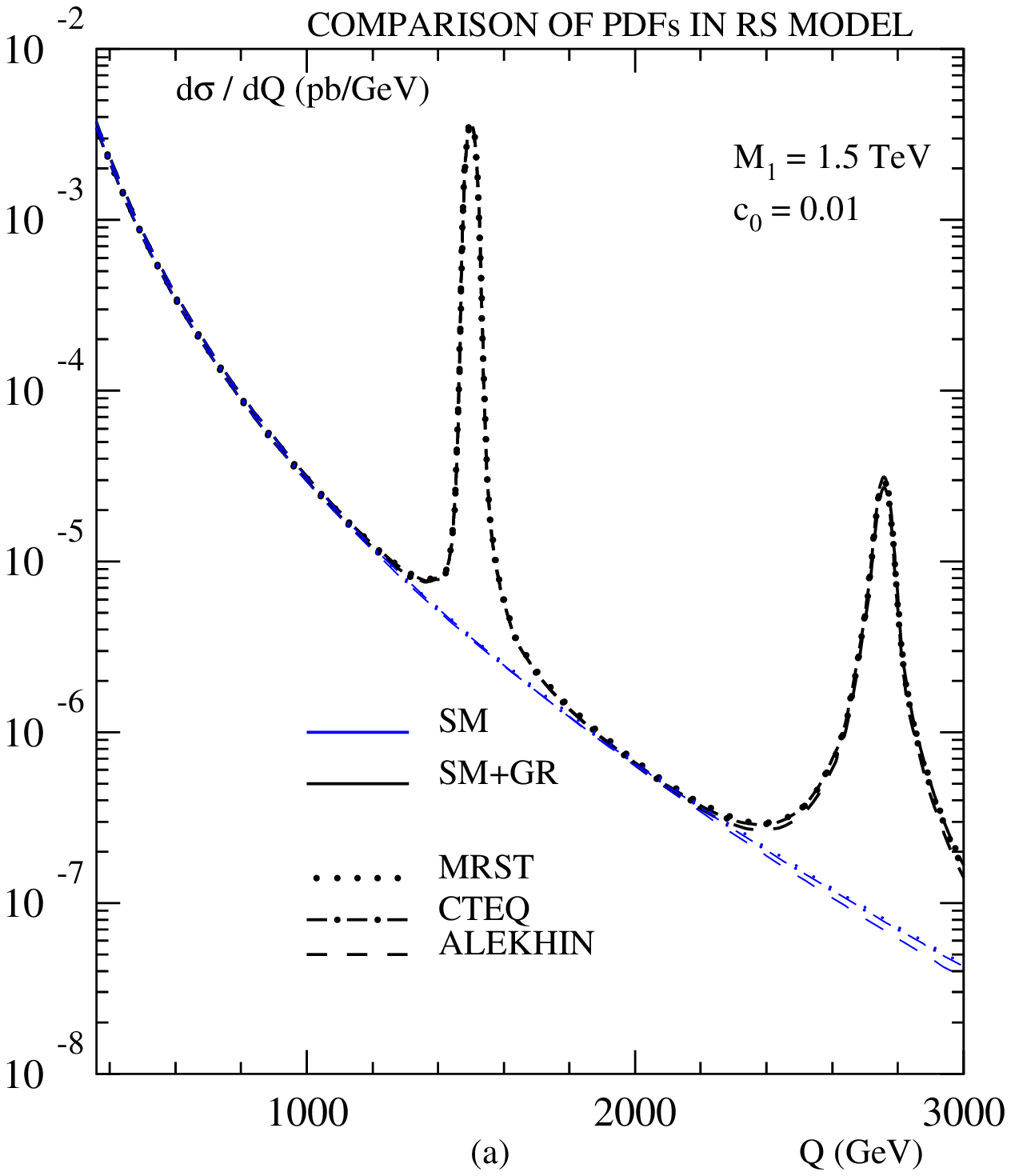,width=8cm,height=10cm,angle=0}
\epsfig{file=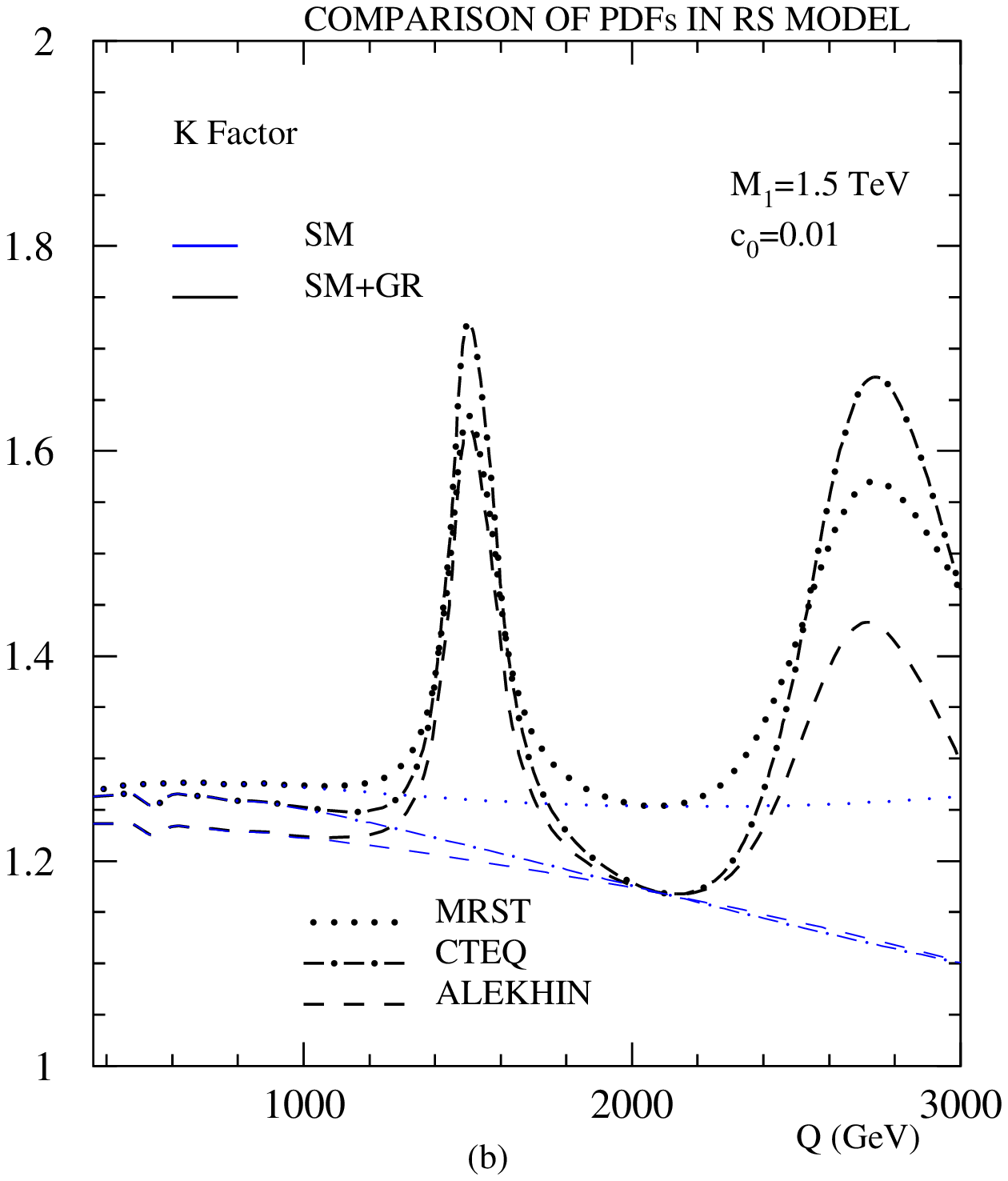,width=8cm,height=10cm,angle=0}
}
\caption{(a) The invariant mass distribution of dilepton pair production at 
LHC in the RS model for various PDfs. (b) The corresponding K-factor for
the various PDFs.}
\label{rsQ}
\end{figure}
\begin{figure}[htb]
\centerline{
\epsfig{file=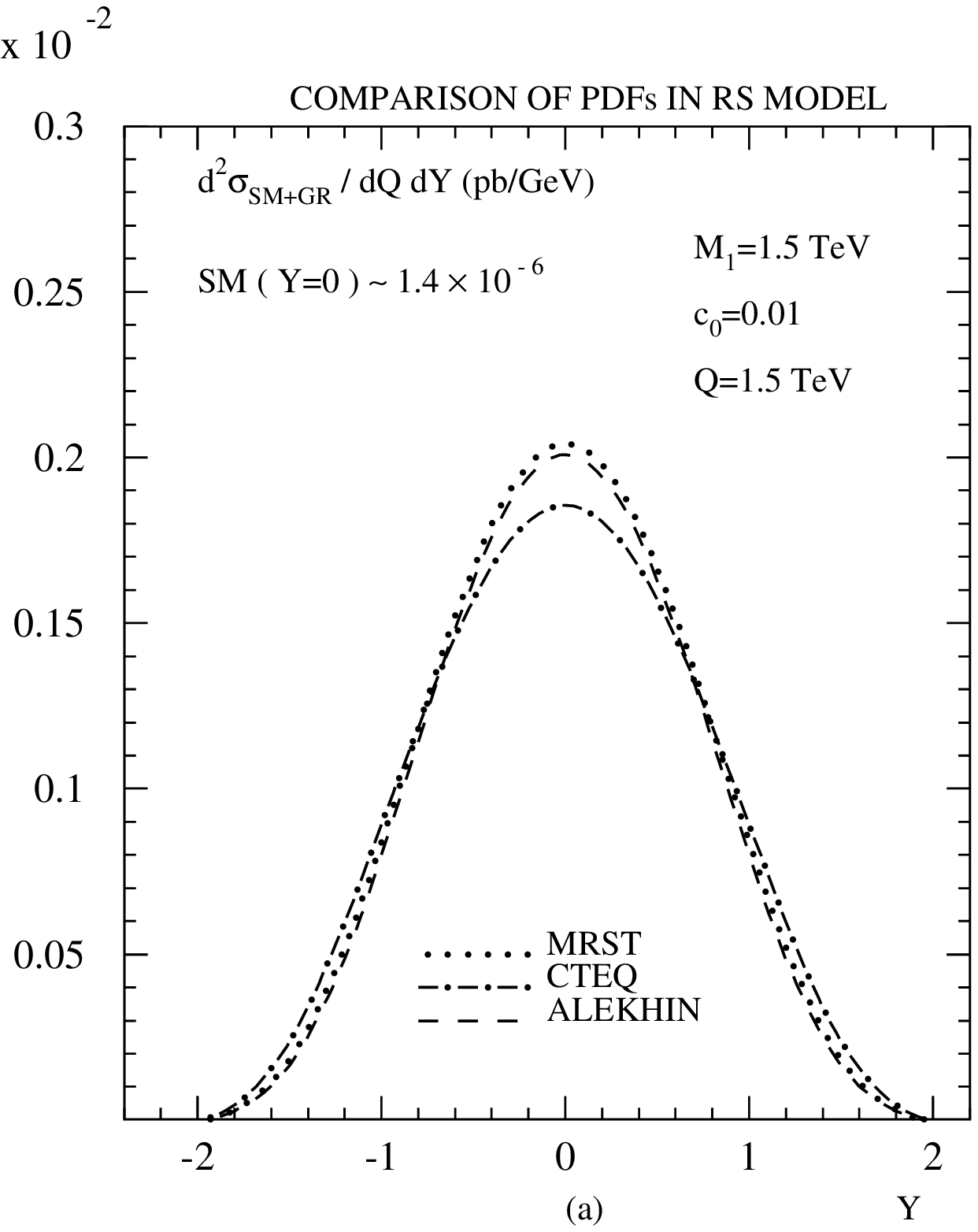,width=8cm,height=9cm,angle=0}
\epsfig{file=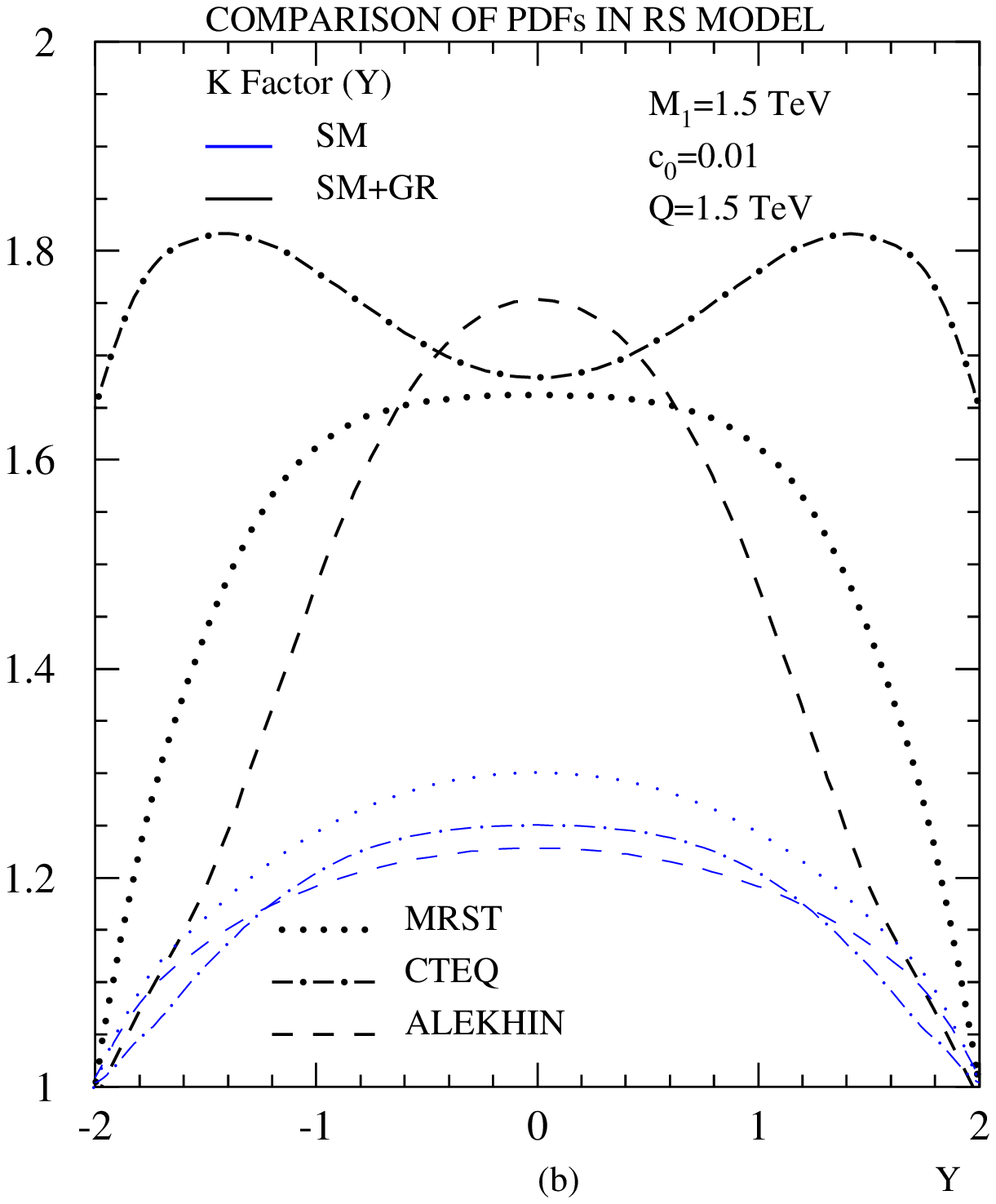,width=8cm,height=9cm,angle=0}
}
\centerline{
\epsfig{file=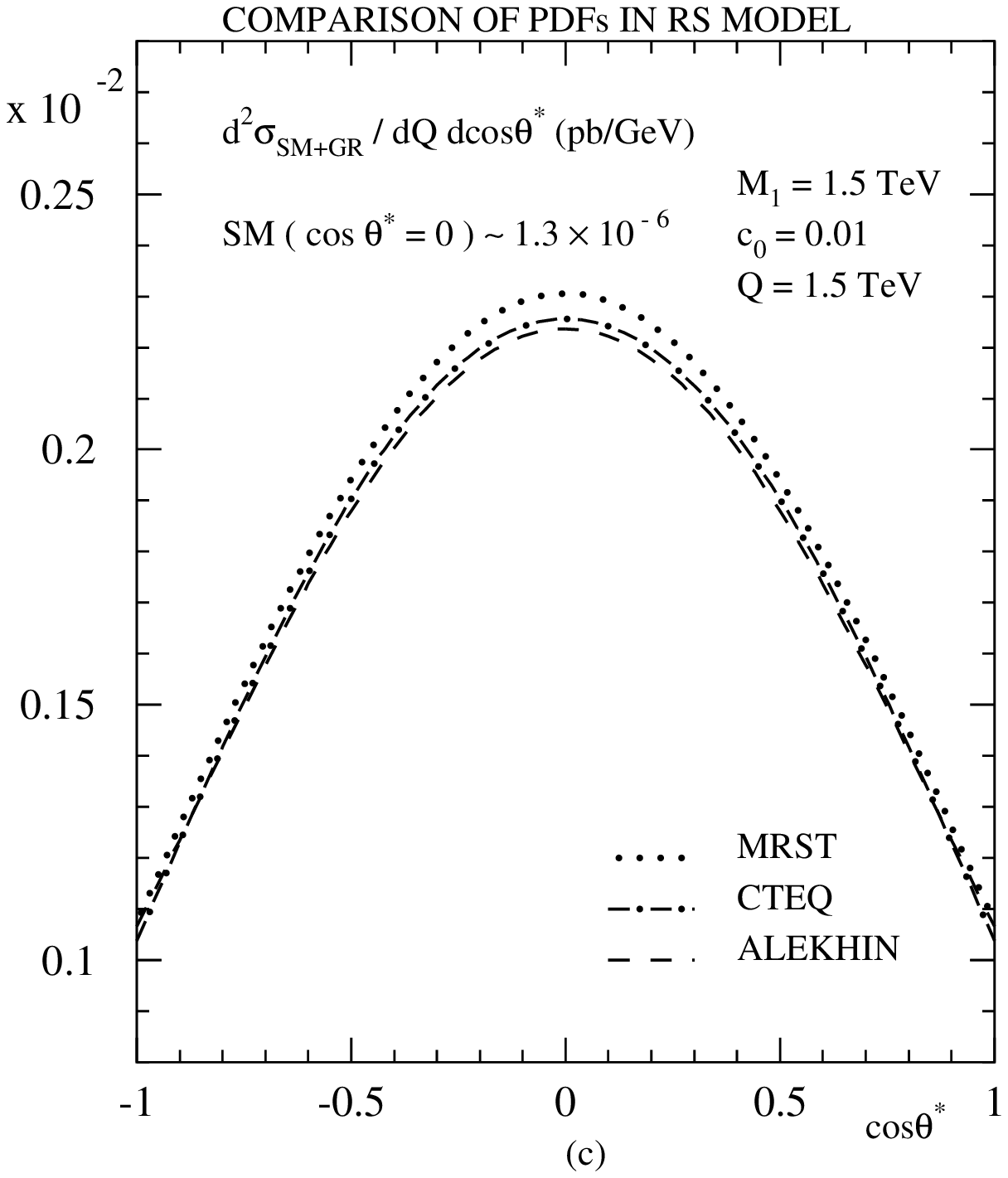,width=8cm,height=9cm,angle=0}
\epsfig{file=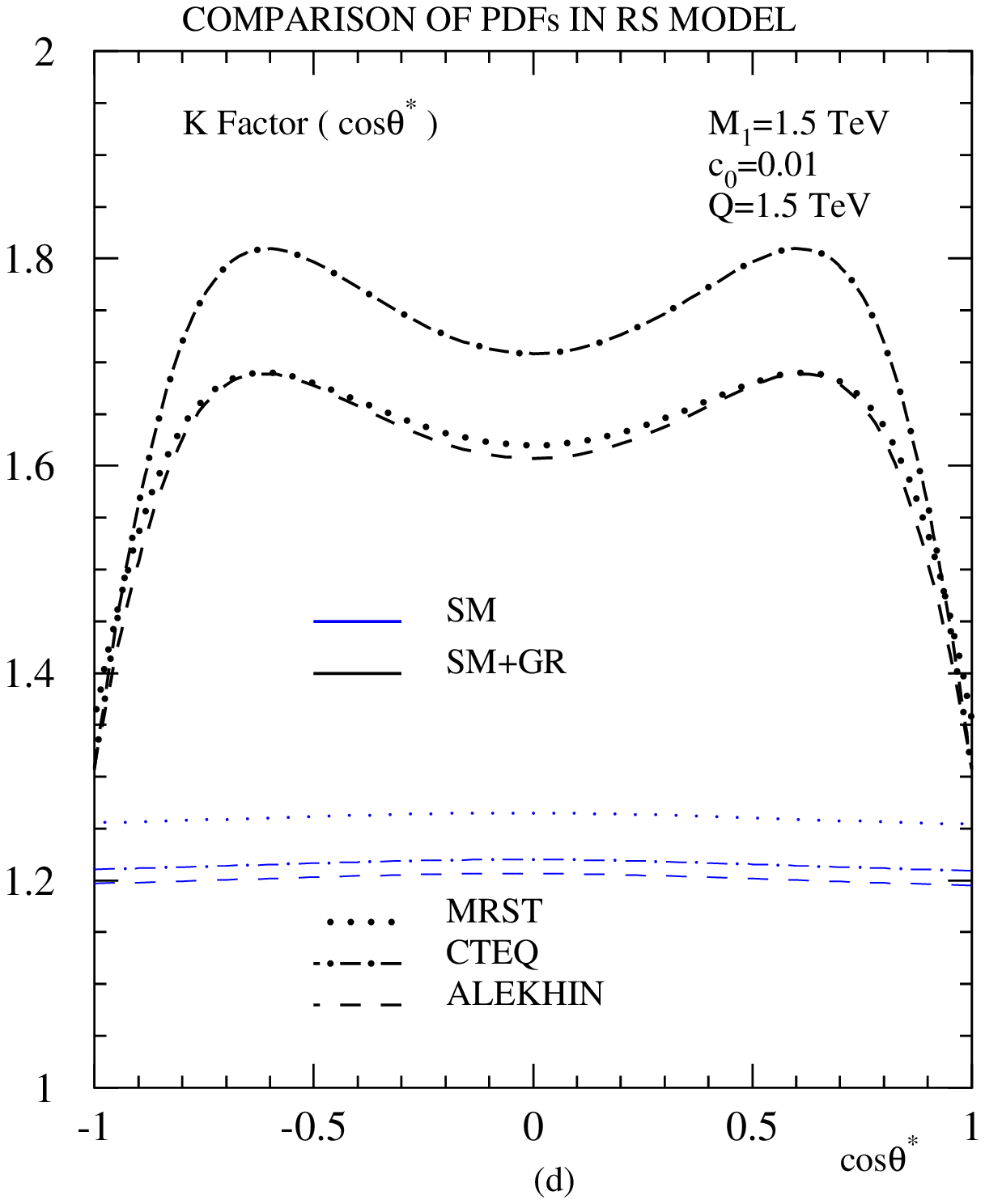,width=8cm,height=9cm,angle=0}
}
\caption{(a) The double differential cross section with respect invariant mass
and rapidity for various PDFs in the RS model at $Q=1.5$ TeV, the region of 
first resonance. (b) The corresponding K-factor as function of rapidity 
at $Q=1.5$ TeV.
(c) In the region of first RS resonance, the double differential
with respect to invariant mass and angular distribution of the lepton is
plotted for the various PDFs at LHC.  (d) The corresponding K-factor for the
various PDFs.
}
\label{rsY}
\end{figure}

\begin{figure}[htb]
\centerline{
\epsfig{file=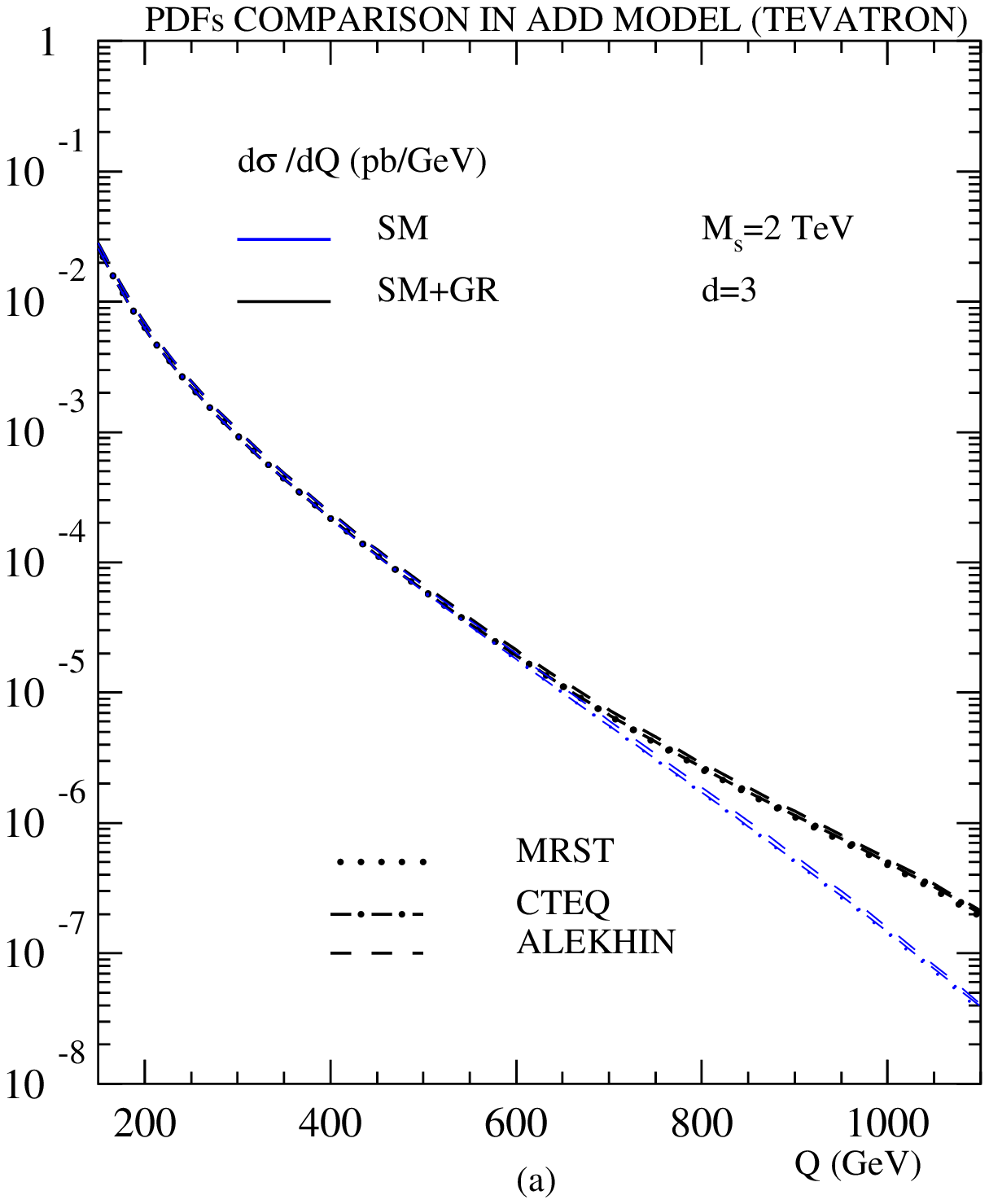,width=8cm,height=9cm,angle=0}
\epsfig{file=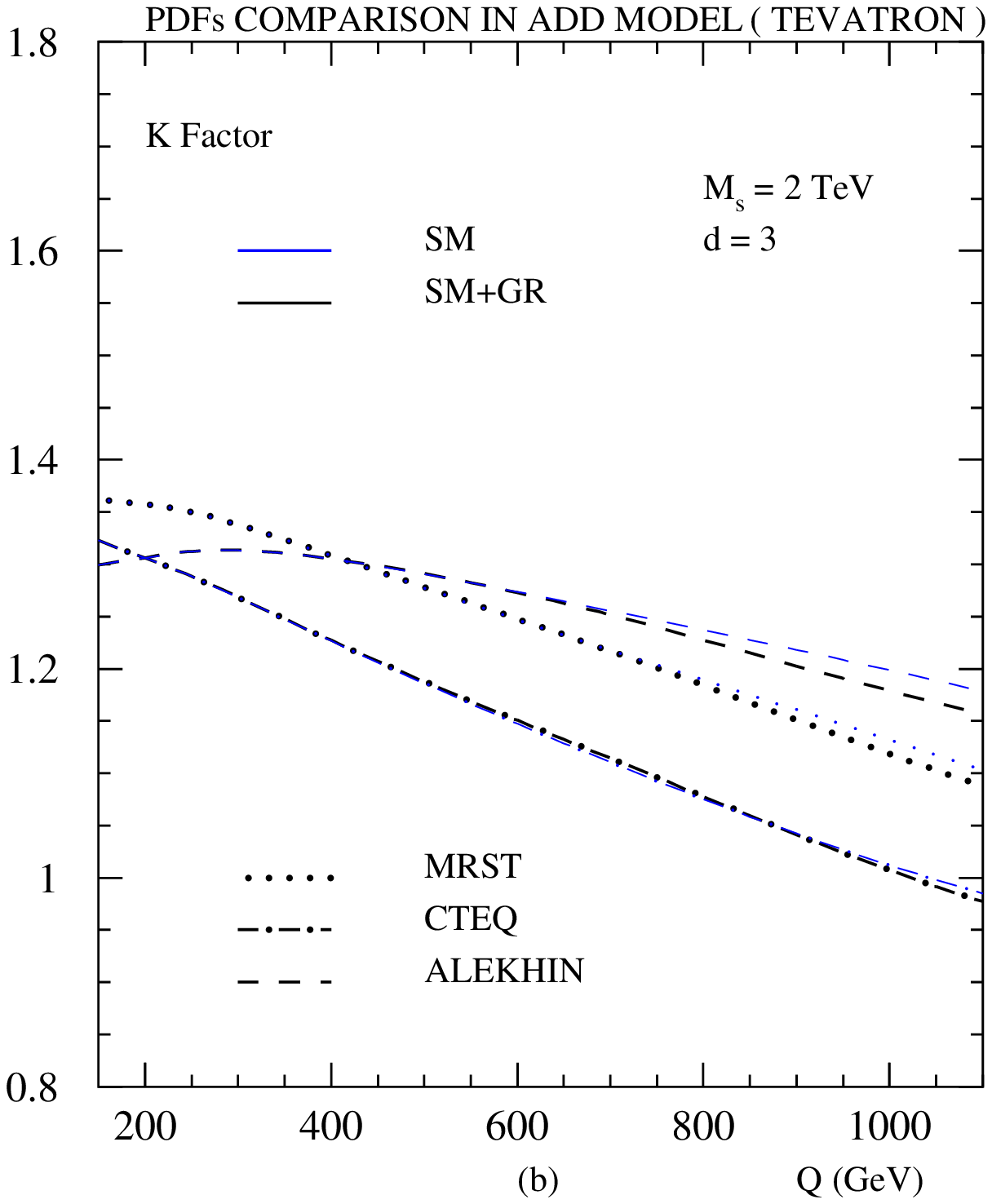,width=8cm,height=9cm,angle=0}
}
\centerline{
\epsfig{file=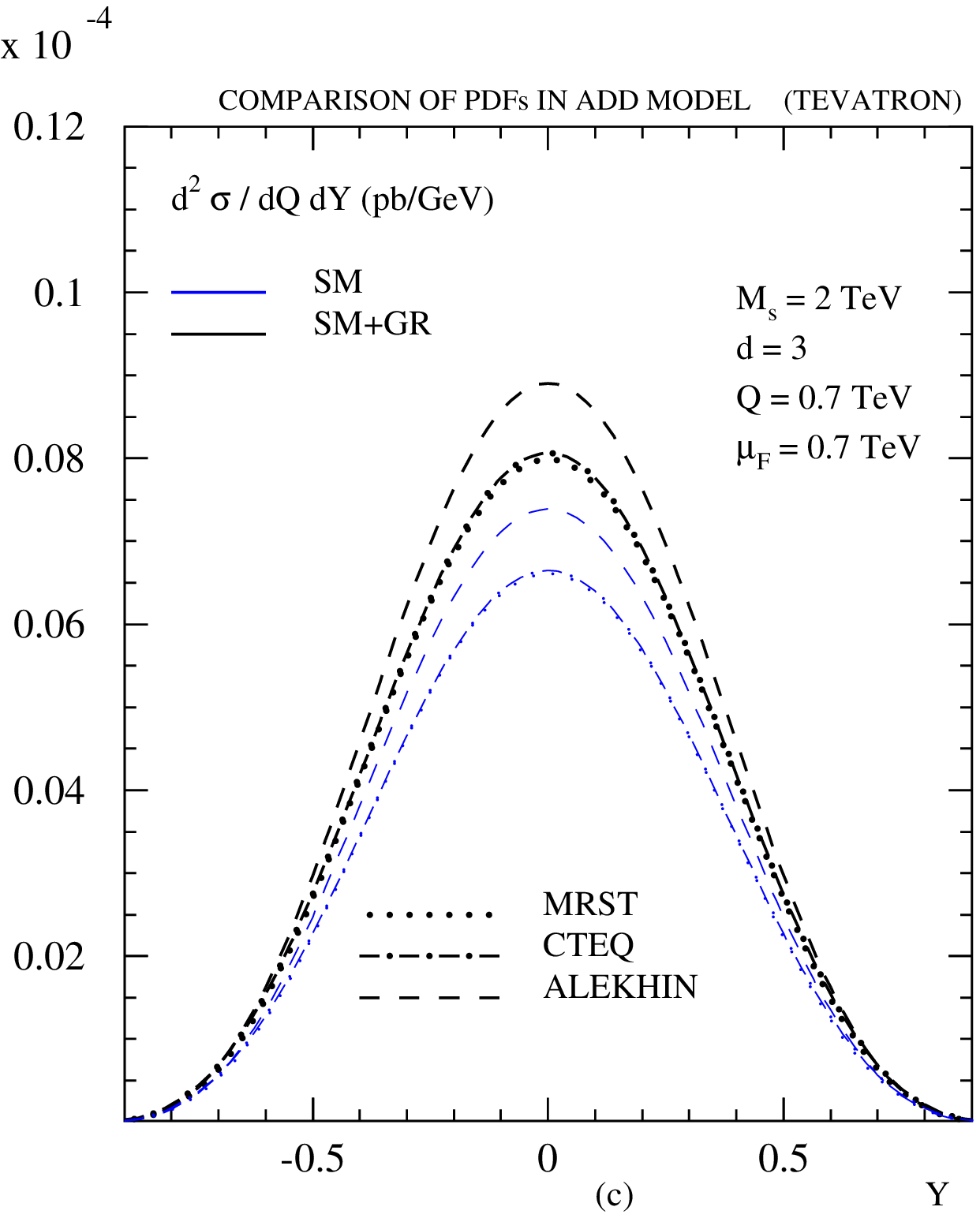,width=8cm,height=9cm,angle=0}
\epsfig{file=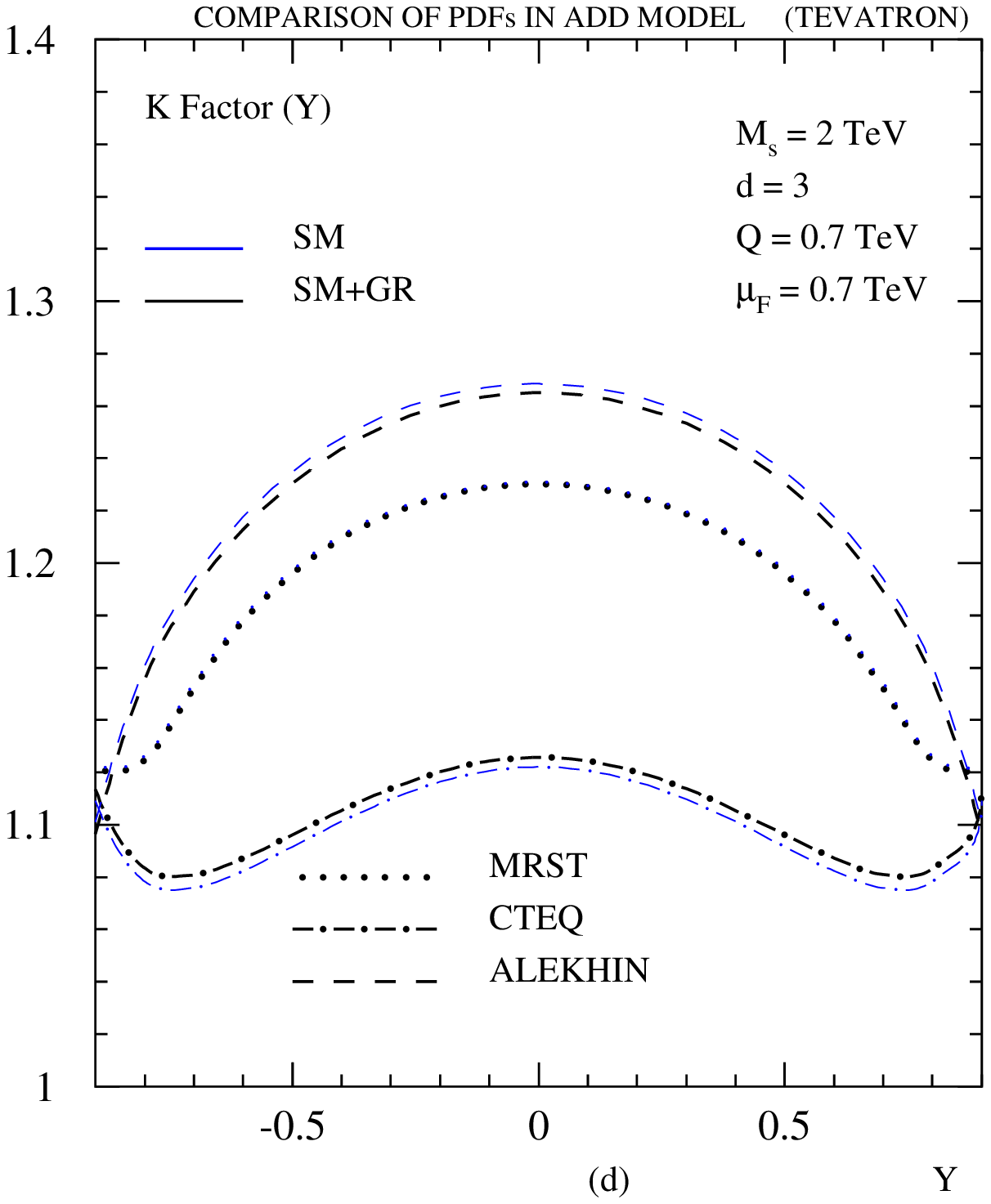,width=8cm,height=9cm,angle=0}
}

\caption{ADD model at Tevatron for the various PDF set, we plot in
(a) the invariant mass distribution. In (b) the corresponding K-factor.
(c) The double differential with respect to $Q$ and $Y$ is plotted for
a fixed $Q=0.7$ TeV and for the $Y$ range of Tevatron.  In (d) the
corresponding K factor is plotted.
}
\label{tev1}
\end{figure}

\begin{figure}[htb]
\centerline{
\epsfig{file=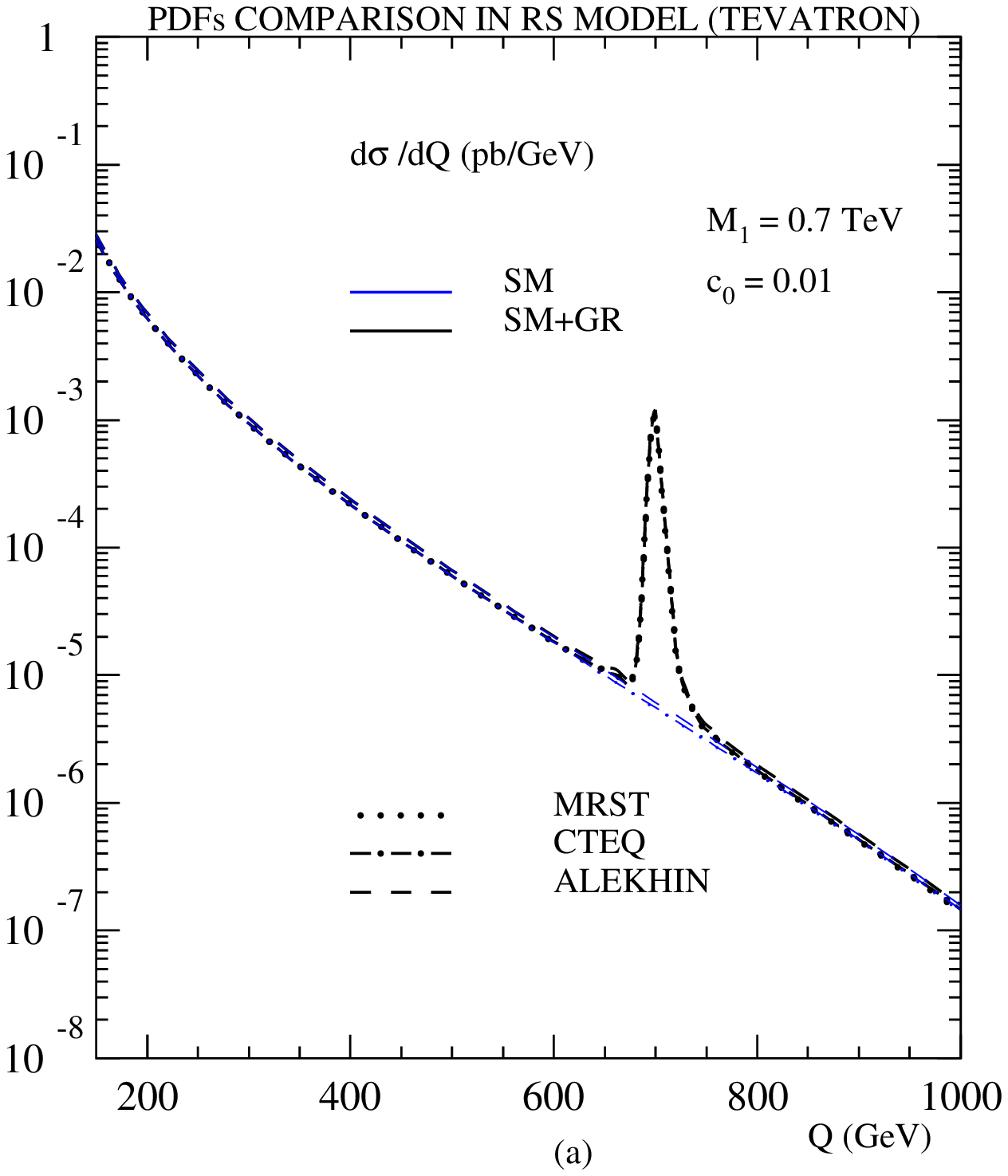,width=8cm,height=9cm,angle=0}
\epsfig{file=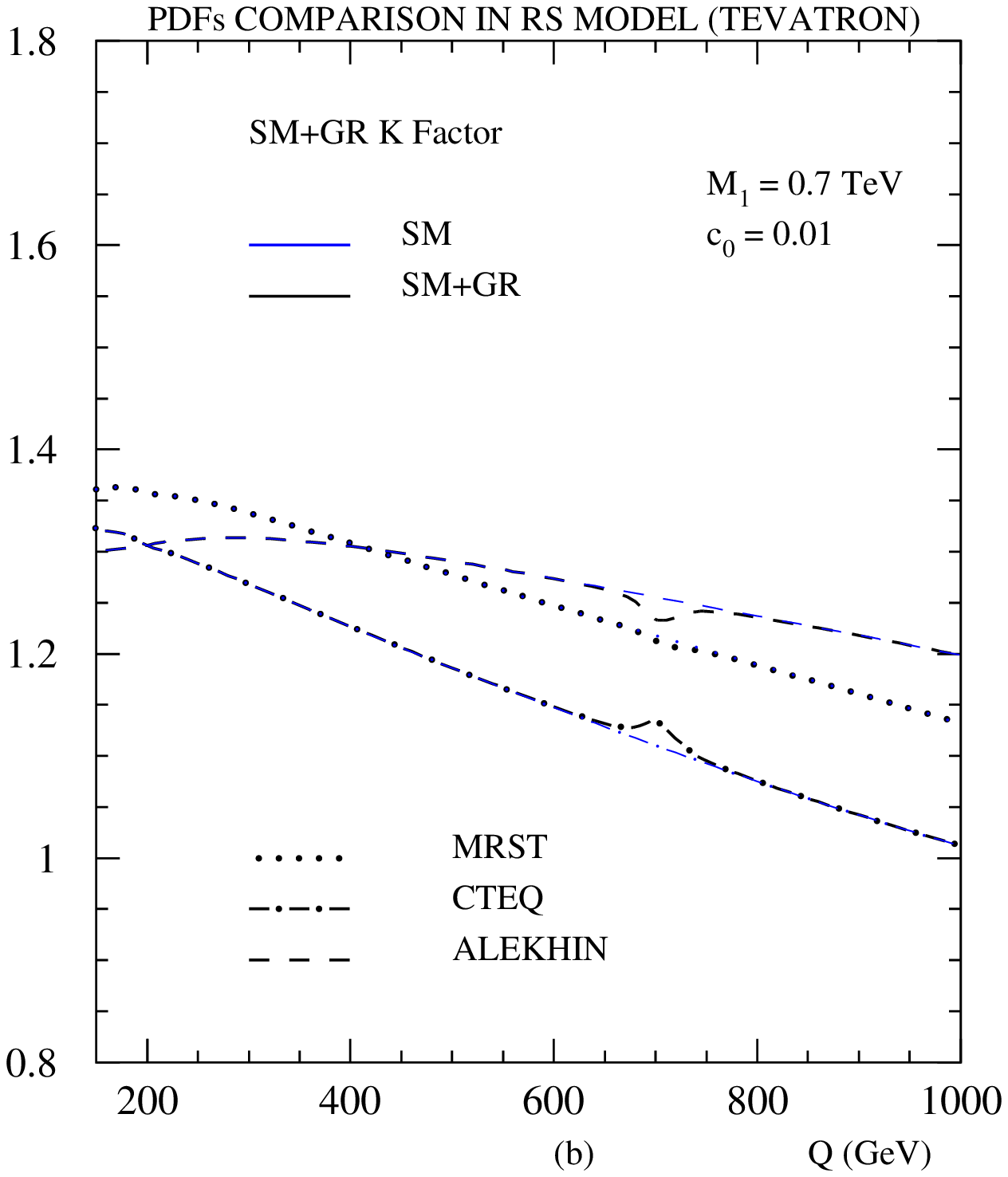,width=8cm,height=9cm,angle=0}
}
\centerline{
\epsfig{file=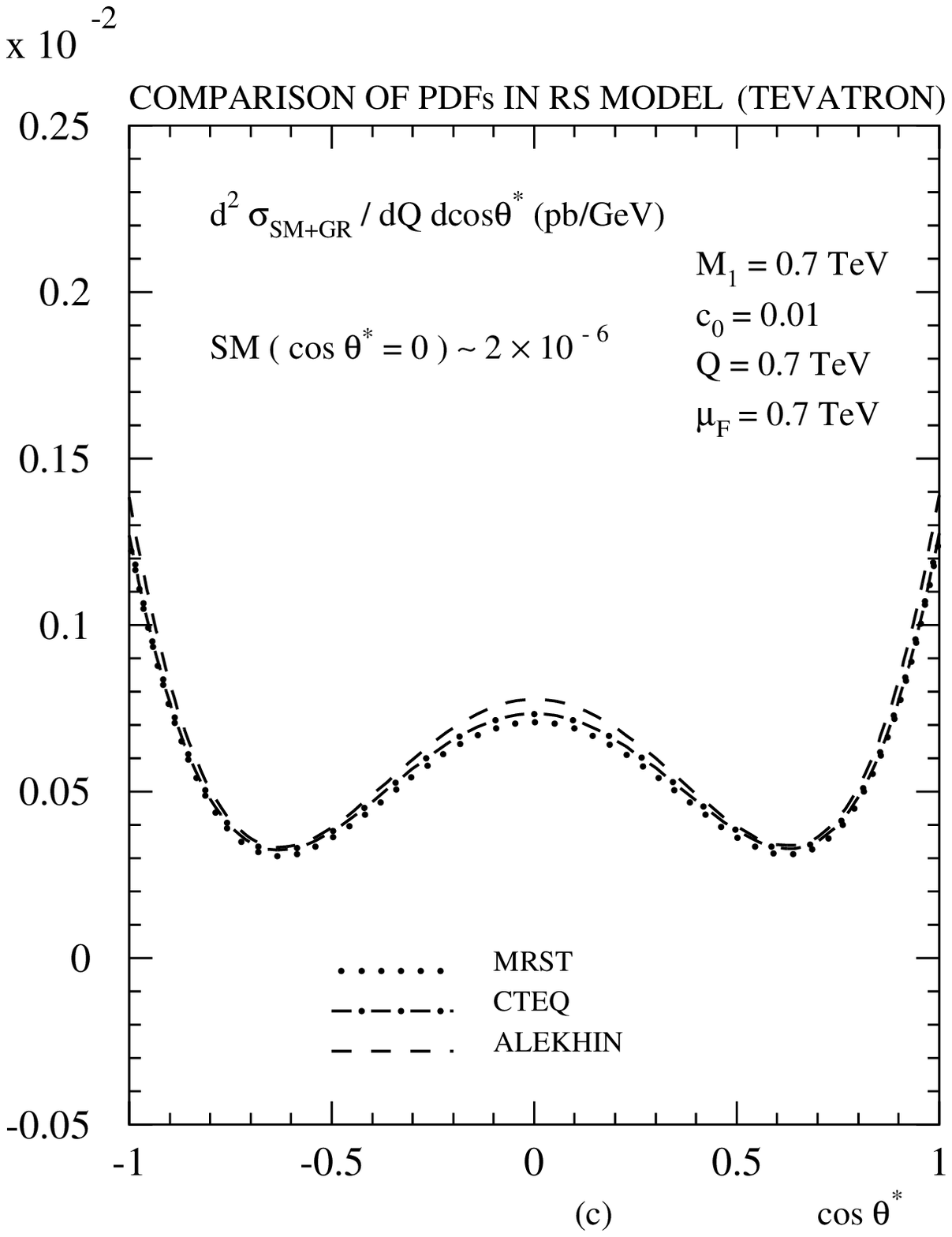,width=8cm,height=9cm,angle=0}
\epsfig{file=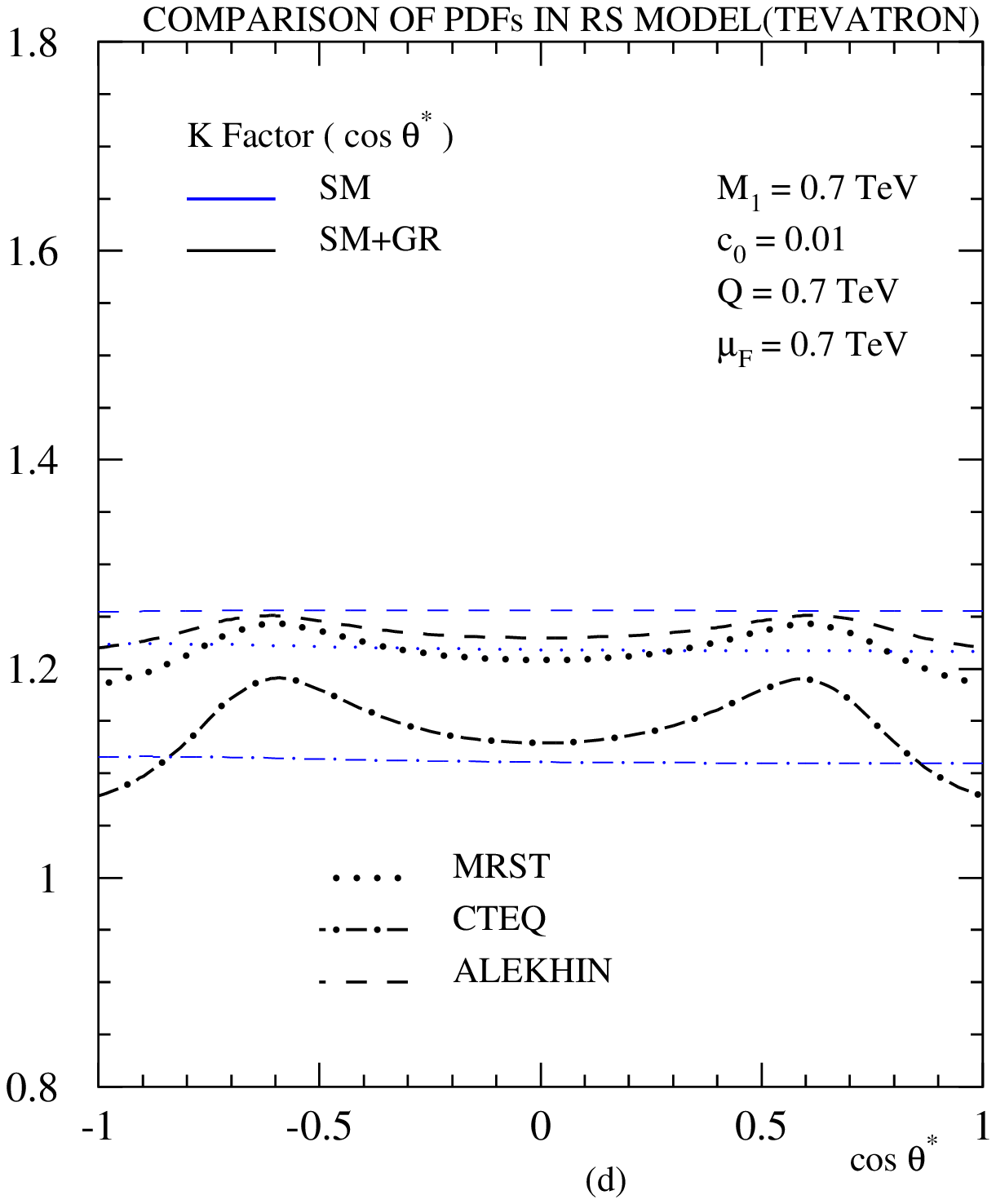,width=8cm,height=9cm,angle=0}
}

\caption{RS model at Tevatron for the various PDF set, we plot in
(a) the invariant mass distribution. In (b) the corresponding K-factor.
(c) The double differential with respect to $Q$ and $\cos \theta^*$ is
plotted for a fixed $Q=0.7$ TeV and for the $\cos \theta^*$.  In (d) the
corresponding K factor is plotted.
}
\label{tev2}
\end{figure}

\begin{figure}[htb]
\centerline{
\epsfig{file=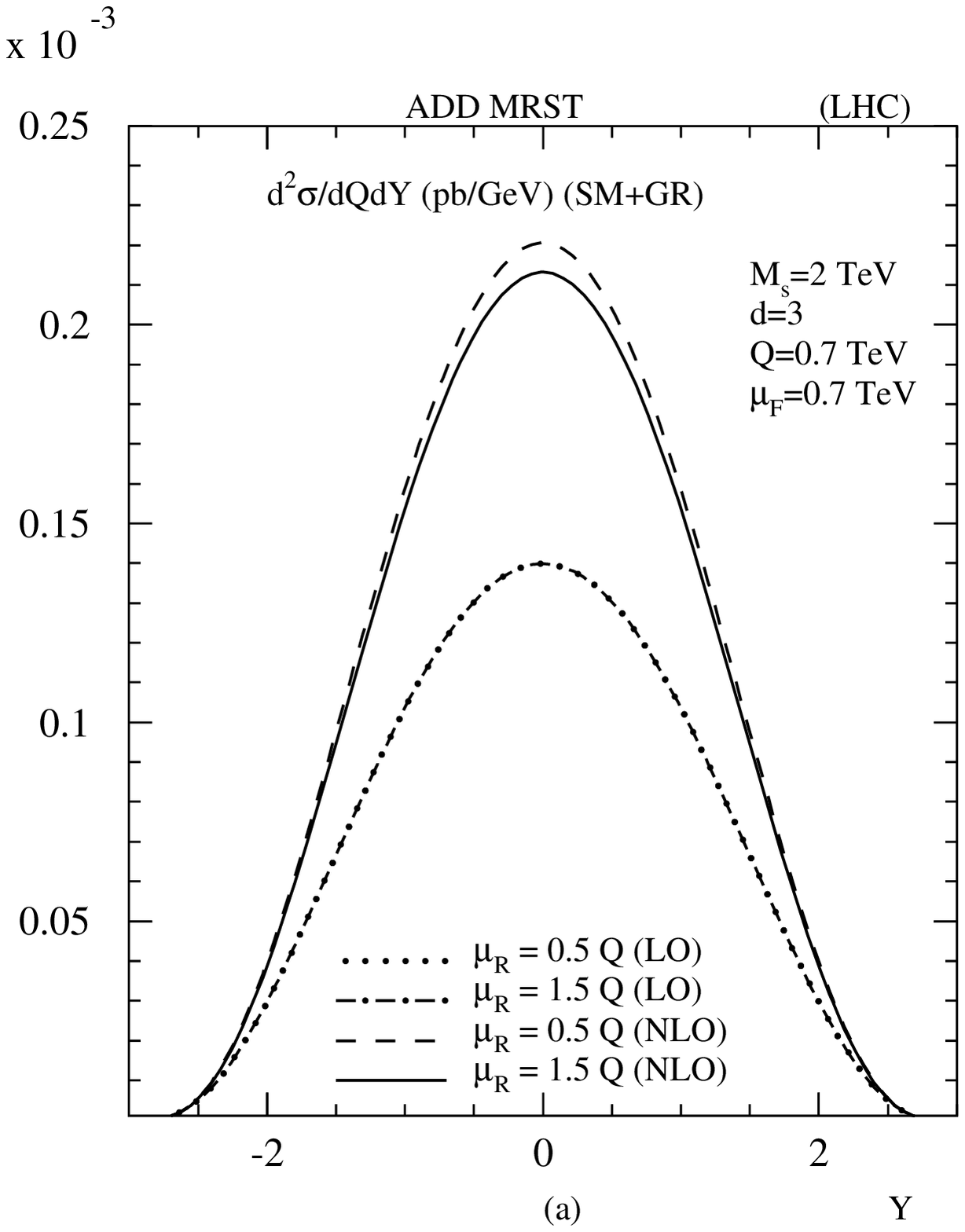,width=8cm,height=10cm,angle=0}
\epsfig{file=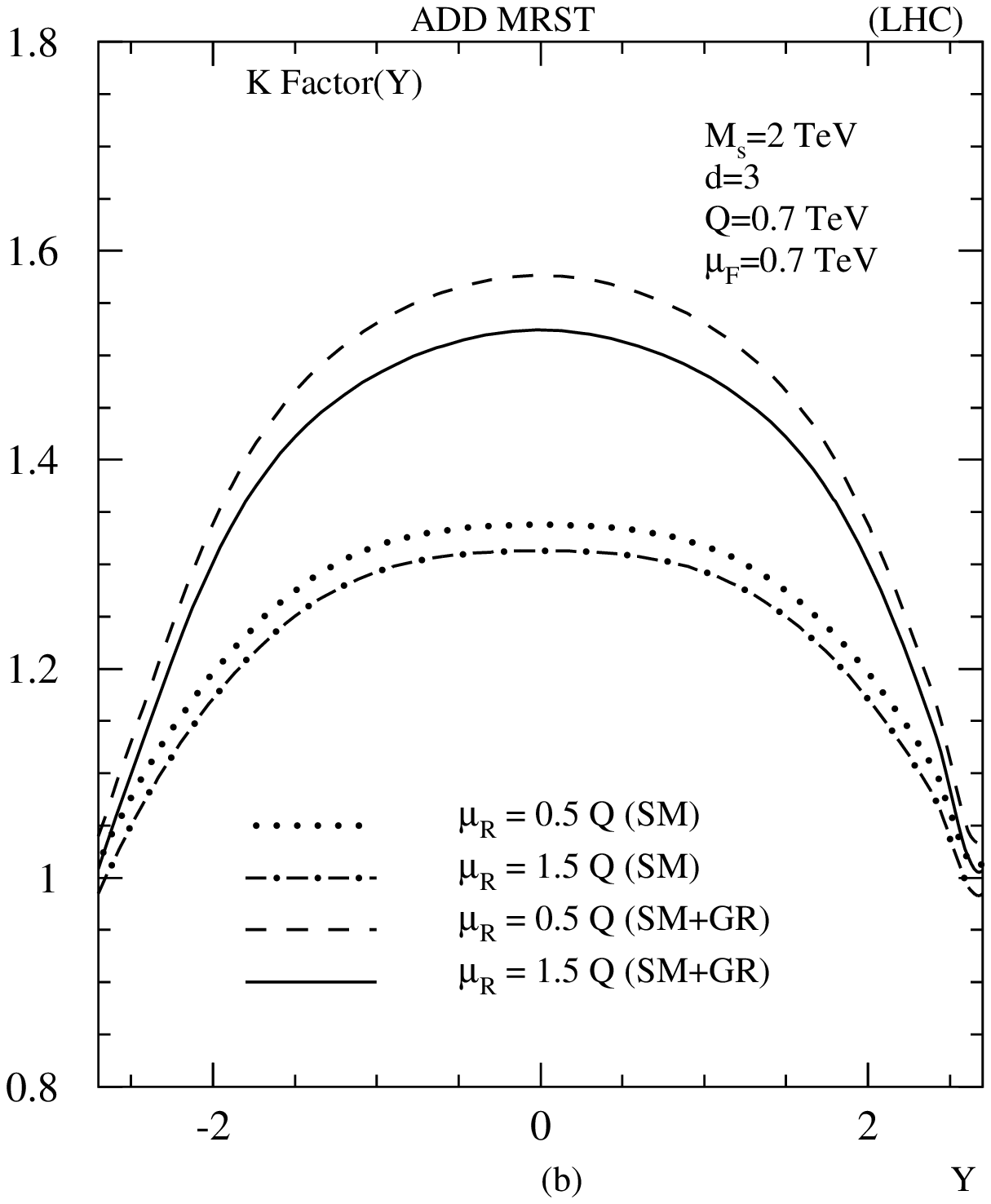,width=8cm,height=10cm,angle=0}
}
\caption{The renormalisation scale dependence for double differential cross 
section as a function of rapidity for $Q=0.7$ TeV.  The PDF used is MRST and
the renormalisation scale is varied in the range $\mu_R=0.5~Q - 1.5~Q$  for LO
and NLO.  (b) The K-factor dependence on $\mu_R$ for both SM and SM+GR.}
\label{reno}
\end{figure}

\begin{figure}[htb]
\centerline{
\epsfig{file=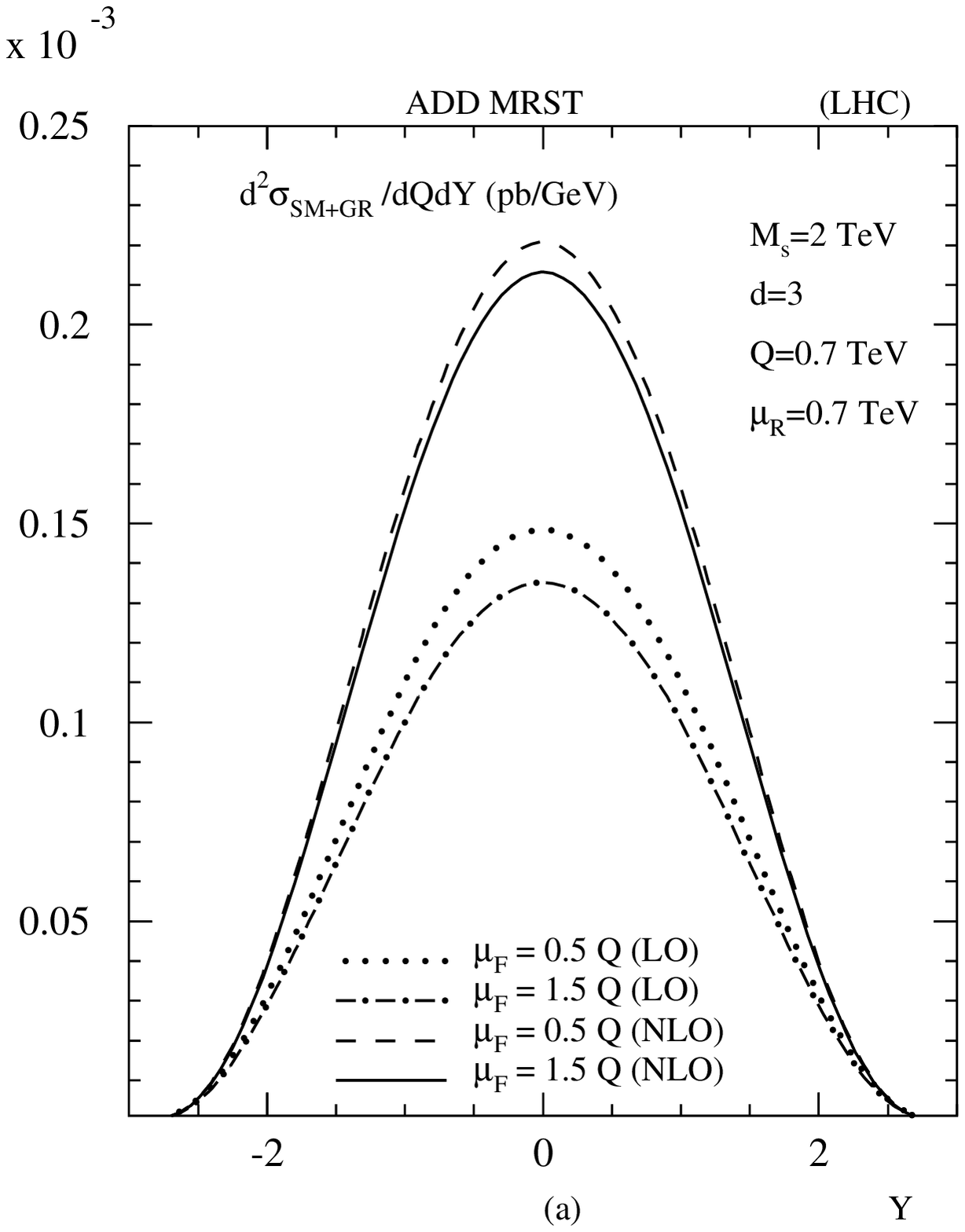,width=8cm,height=9cm,angle=0}
\epsfig{file=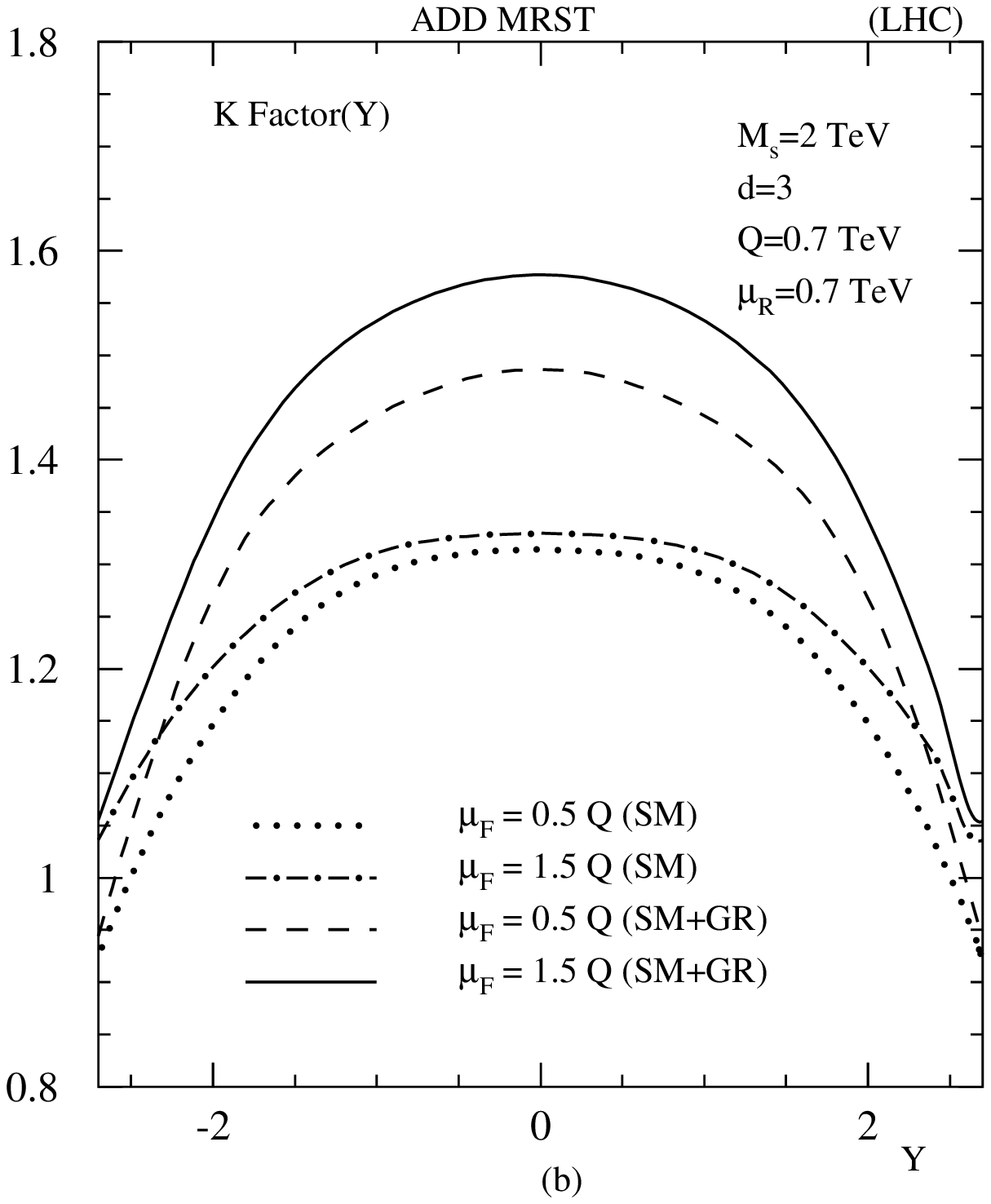,width=8cm,height=9cm,angle=0}
}
\centerline{
\epsfig{file=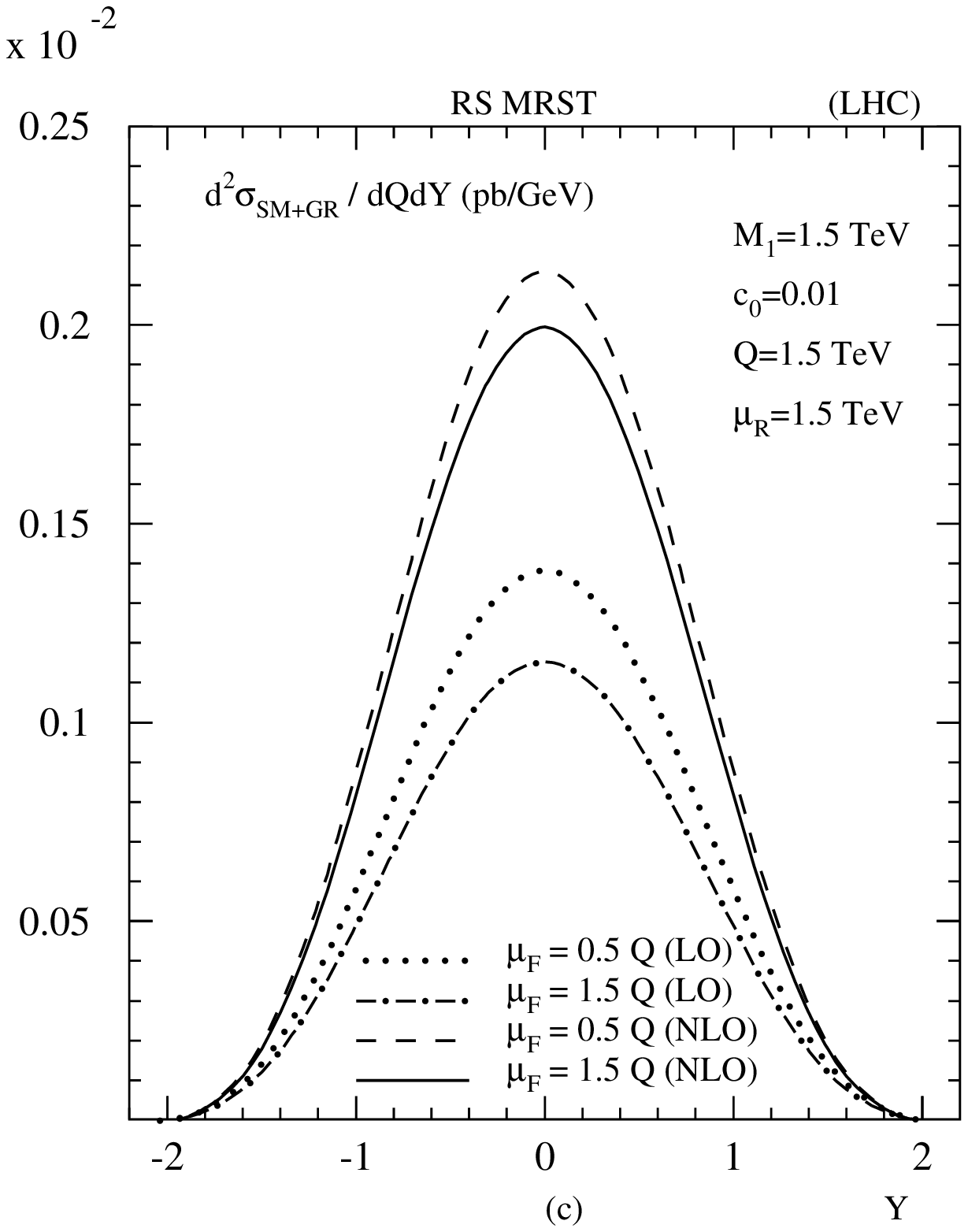,width=8cm,height=9cm,angle=0}
\epsfig{file=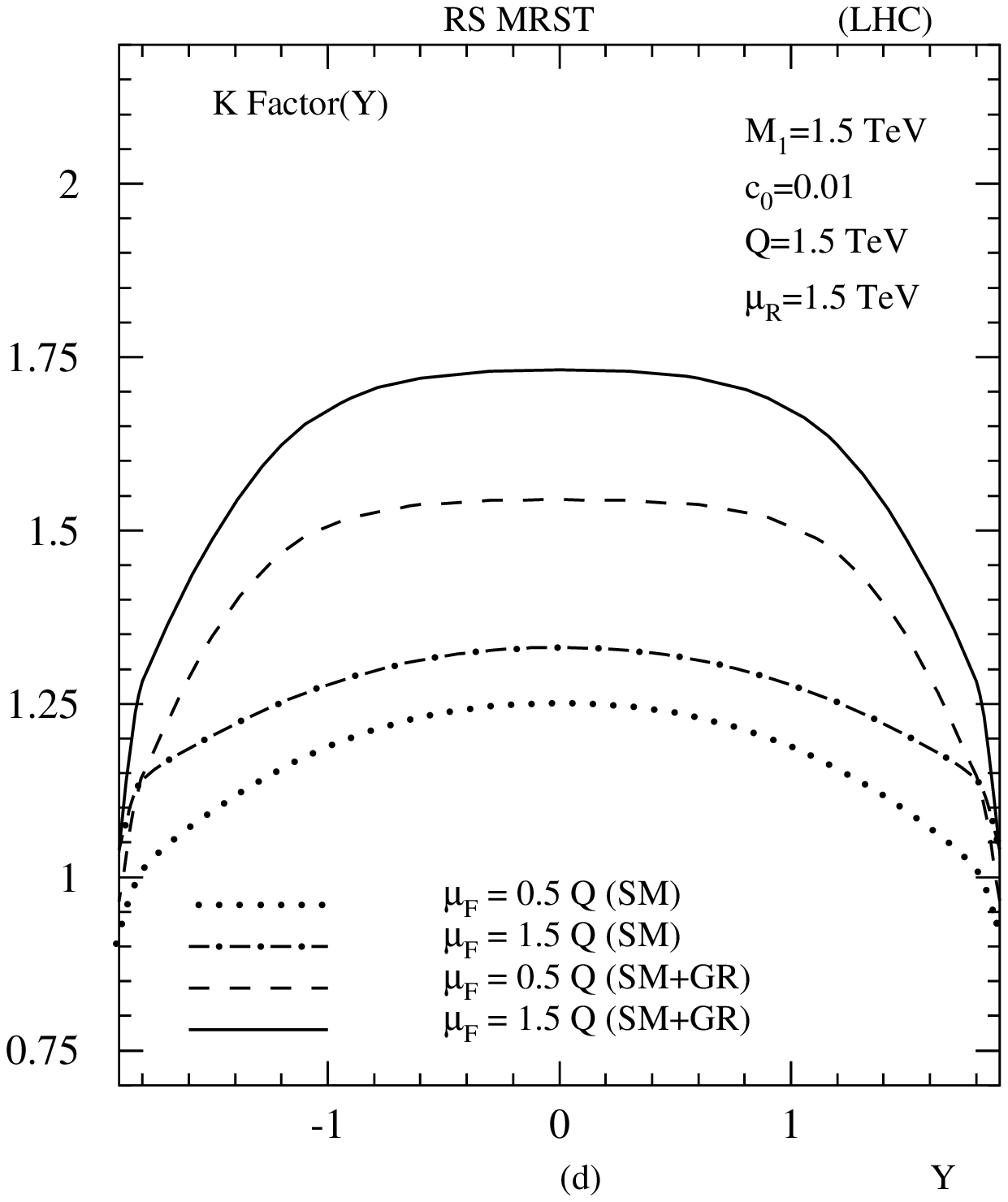,width=8cm,height=9cm,angle=0}
}
\caption{(a) Factorisation scale dependence for the double differential 
cross section as a function of rapidity for LO and NLO for factorisation 
scale in the range $\mu_F=0.5~Q - 1.5~Q$.  
(b) SM and SM+GR K factor for ADD rapidity distribution in the same 
variation of $\mu_F$.  In (c) 
the RS distribution at $Q=\mu_R=1.5$ TeV in the region of first resonance.
(d) The SM and SM+GR K factor for RS rapidity distribution. 
}
\label{fact_Y}
\end{figure}

\begin{figure}[htb]
\centerline{
\epsfig{file=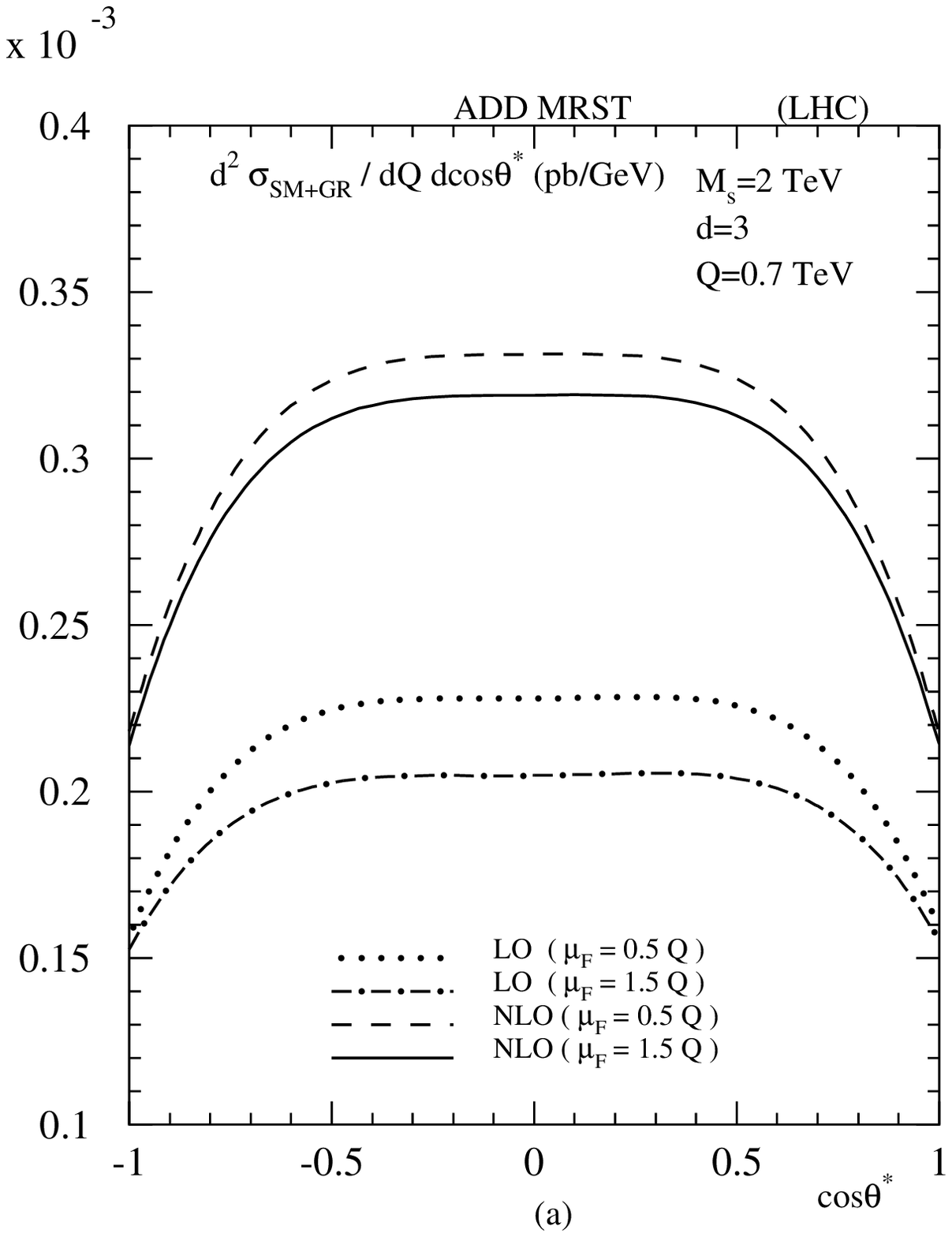,width=8cm,height=9cm,angle=0}
\epsfig{file=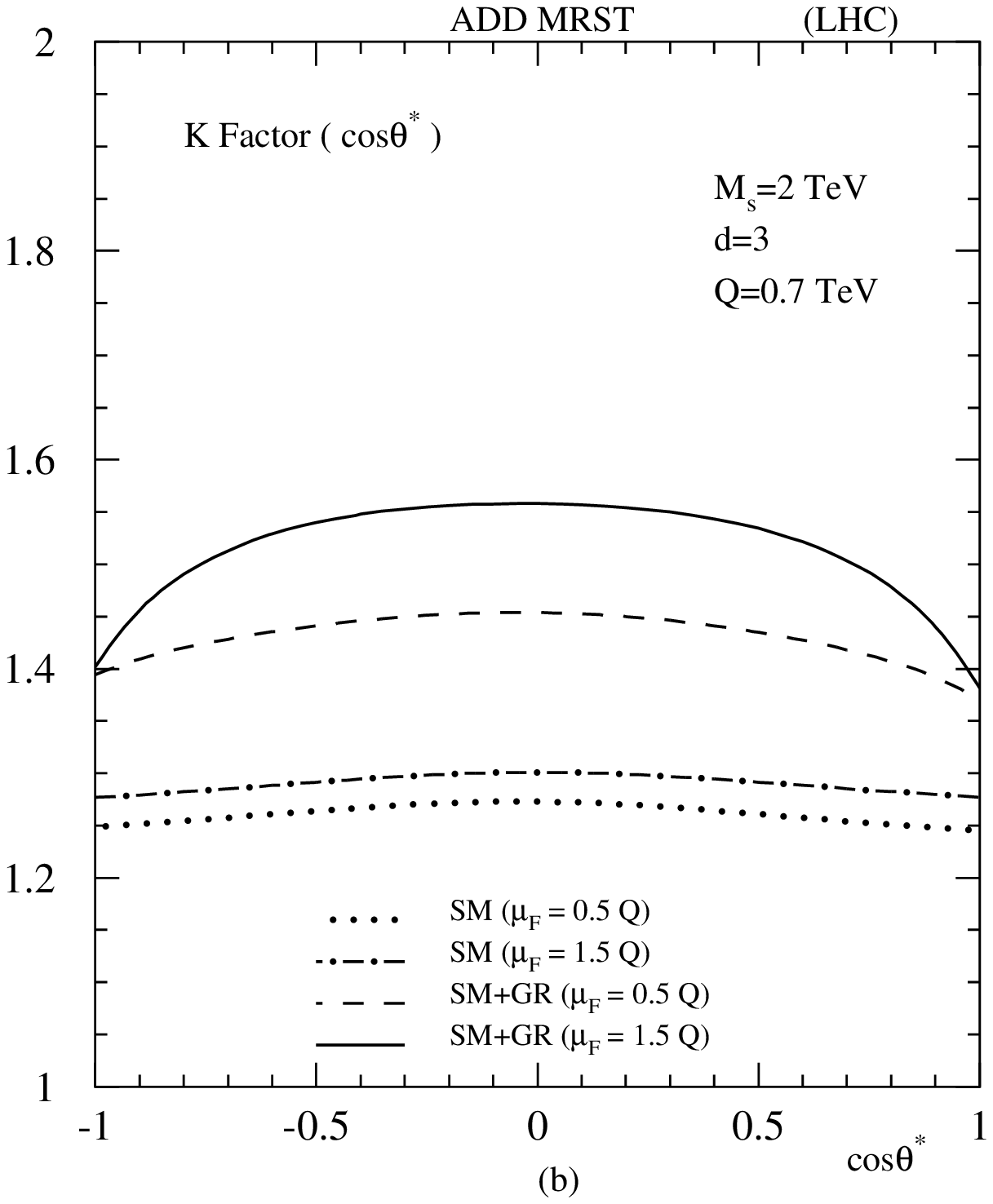,width=8cm,height=9cm,angle=0}
}
\centerline{
\epsfig{file=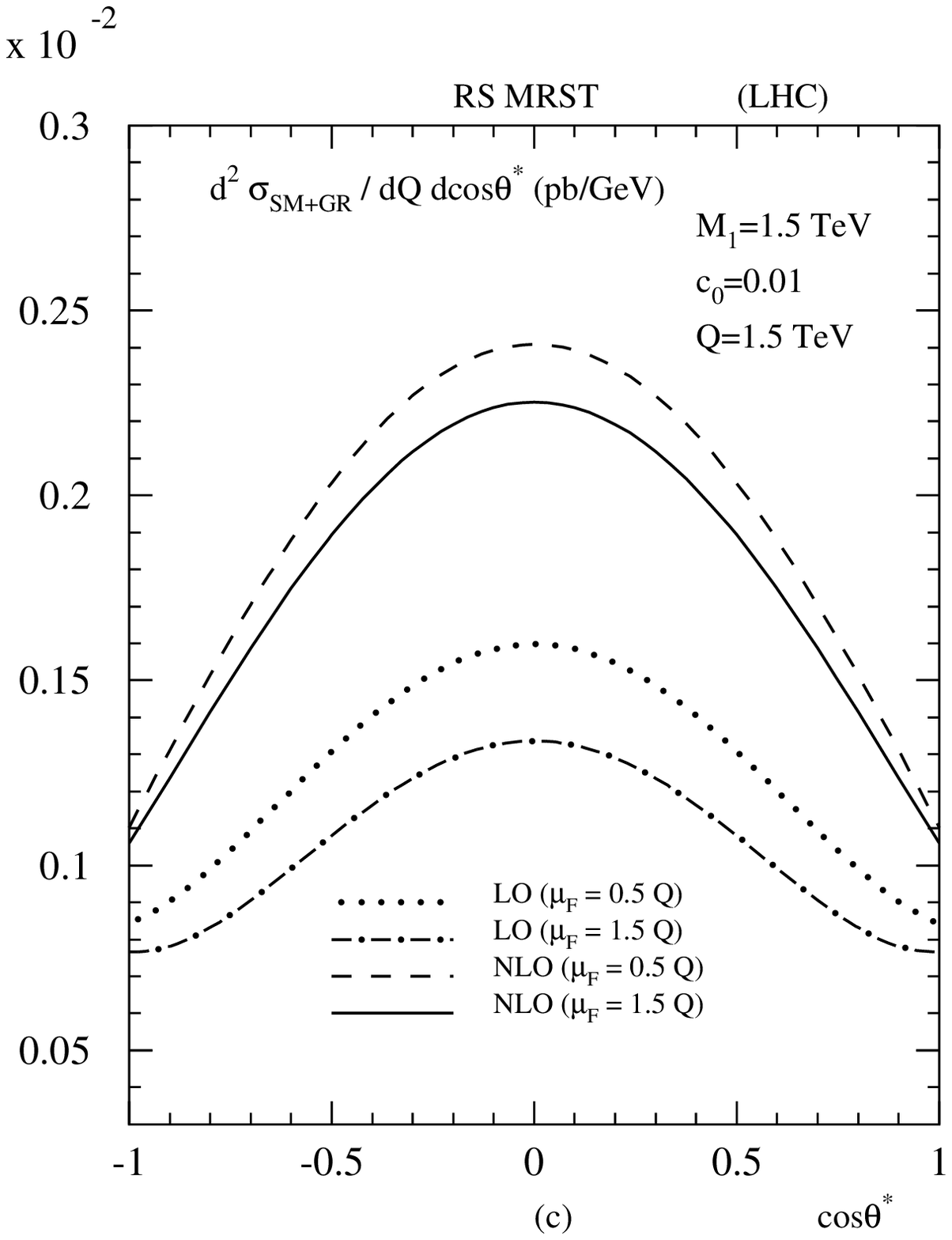,width=8cm,height=9cm,angle=0}
\epsfig{file=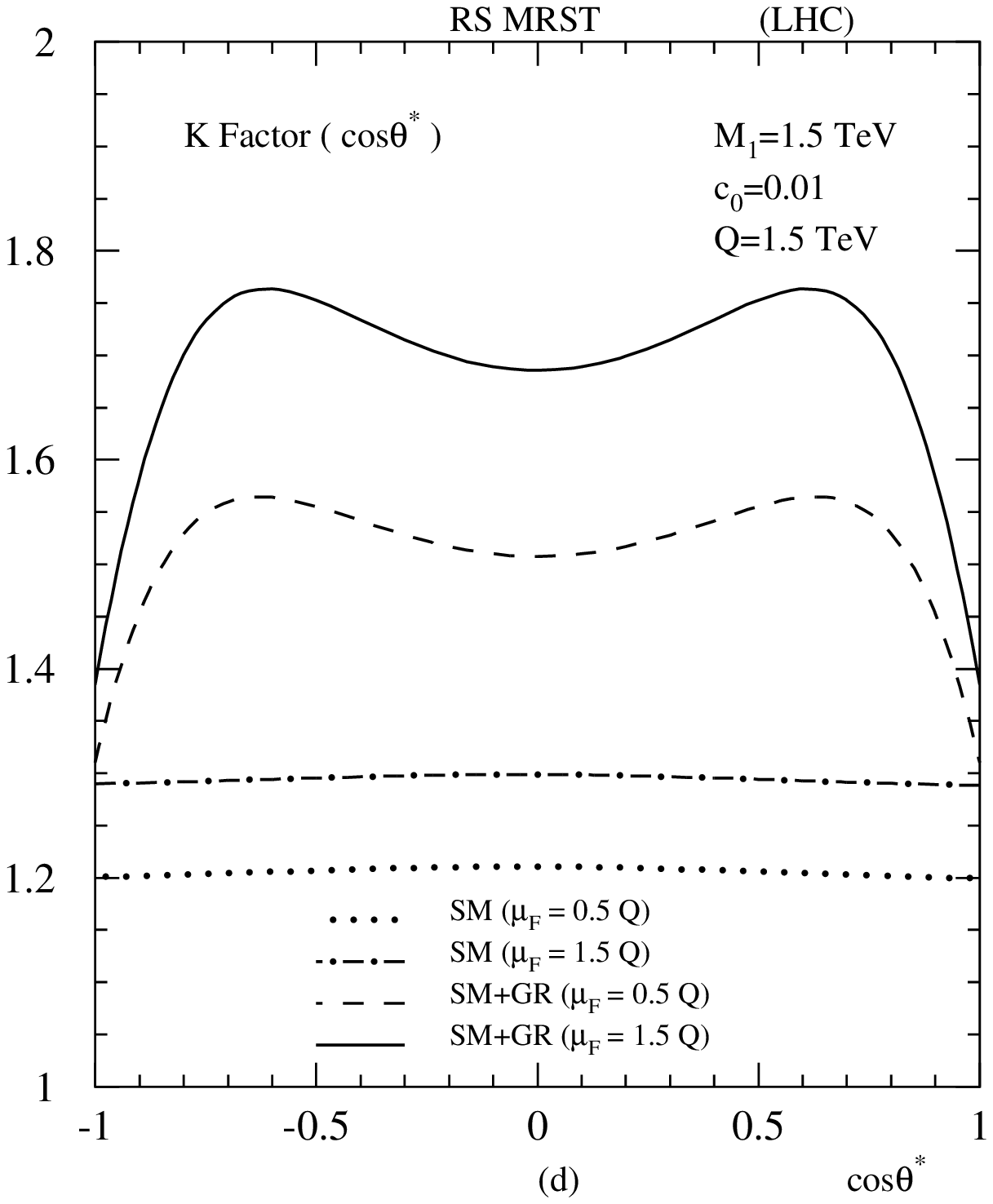,width=8cm,height=9cm,angle=0}
}
\caption{(a) Factorisation scale dependence for the double differential
cross section as a function of $\cos \theta^*$ for LO and NLO for
factorisation scale in the range $\mu_F=0.5~Q - 1.5~Q$.  In (b) we have 
plotted the SM and SM+GR K factor for ADD at $Q=\mu_R=0.7$ TeV.  In (c) 
the RS $\cos \theta^*$ distribution for LO and NLO in the same range 
of $\mu_F$.  (d) The SM and SM+GR K-factor at $Q=\mu_R=1.5$ TeV, the 
region of first resonance.}
\label{fact_cos}
\end{figure}

\begin{figure}[htb]
\centerline{
\epsfig{file=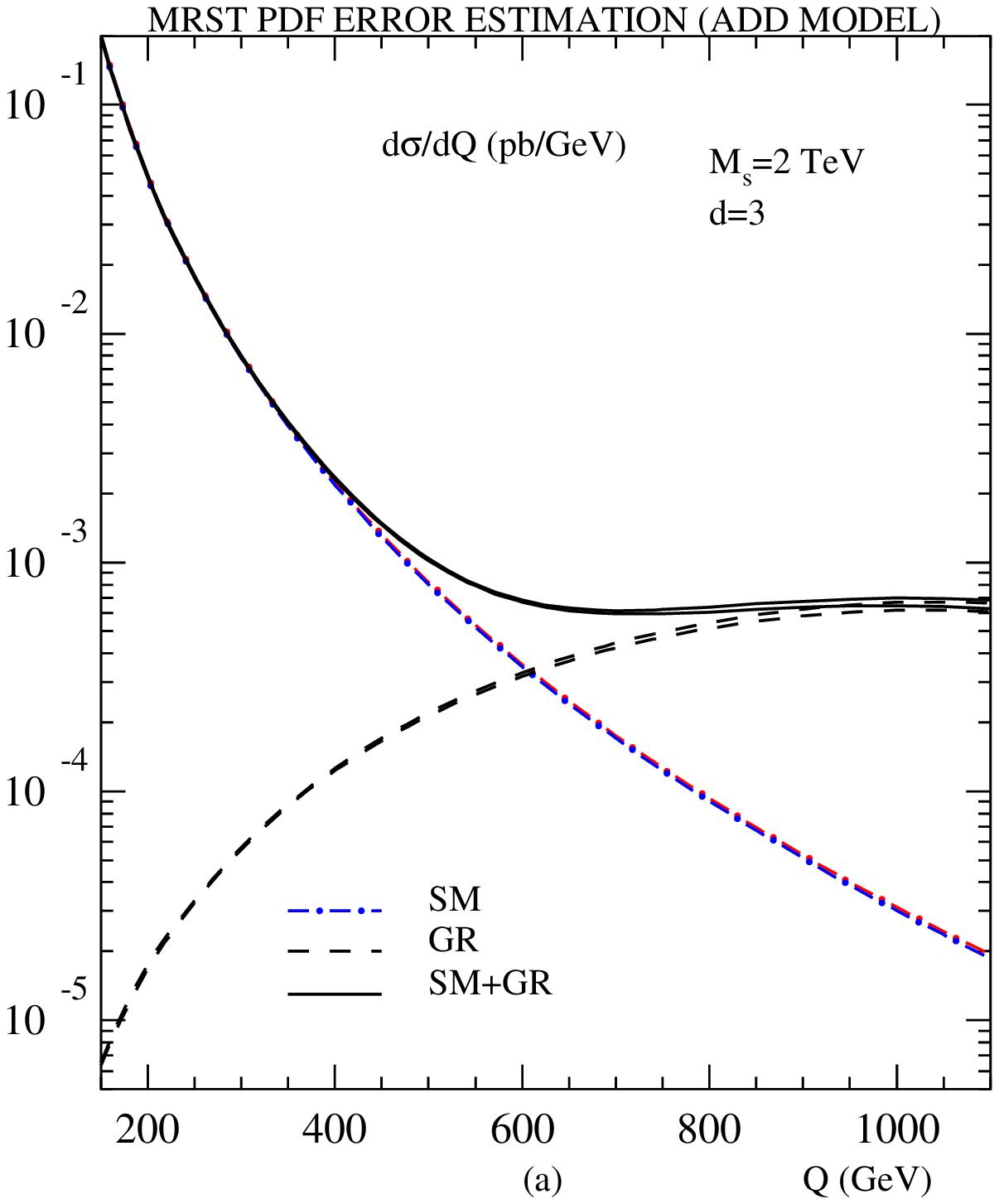,width=8cm,height=10cm,angle=0}
\epsfig{file=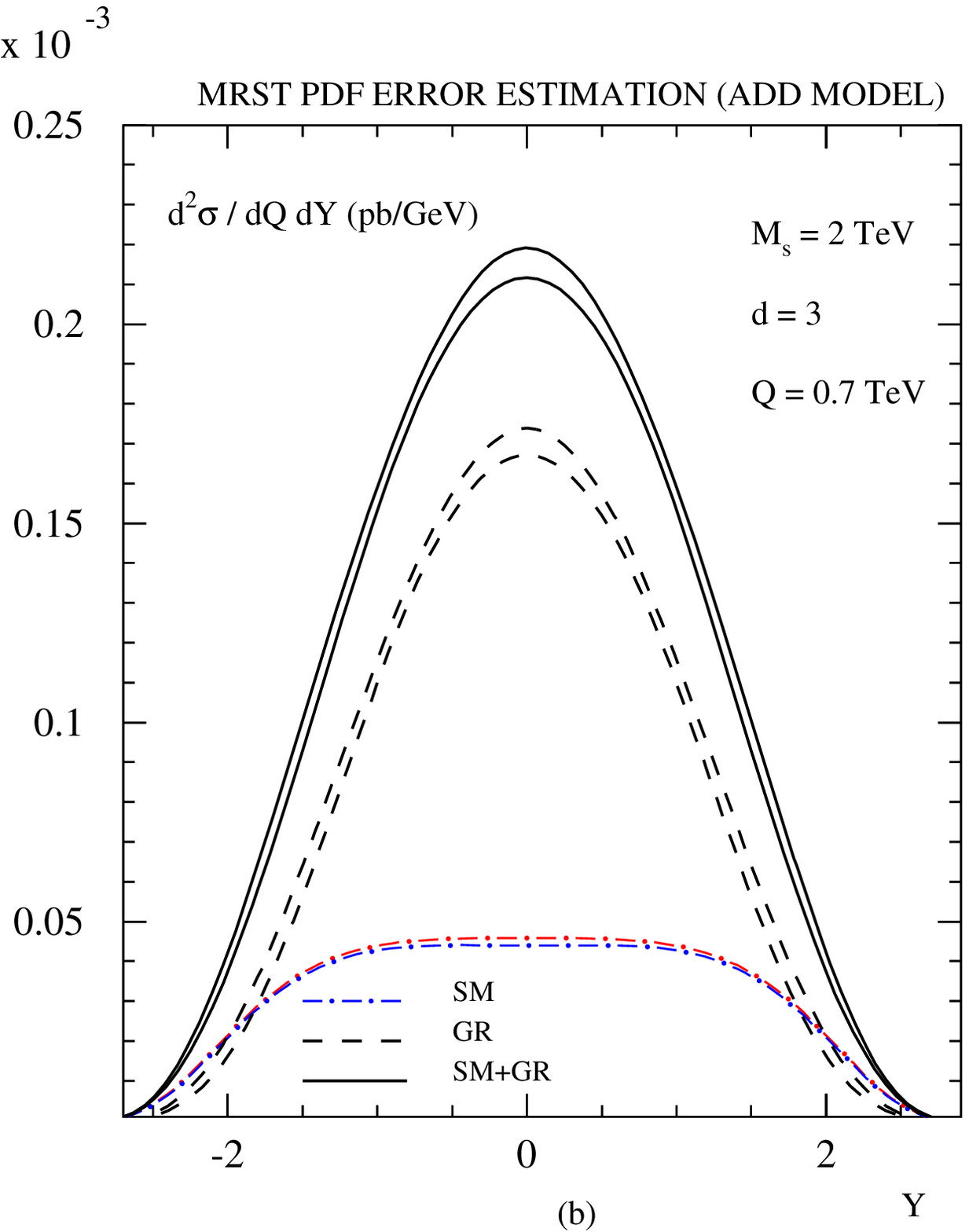,width=8cm,height=10cm,angle=0}
}
\caption{The experimental error on the MRST PDF at LHC in the ADD model for
(a) The invariant mass distribution and (b) The rapidity distribution for a 
fixed $Q=0.7$ TeV.}
\label{err}
\end{figure}


\end{document}